\documentclass[aps,prl,preprint,superscriptaddress]{revtex4-2}
\usepackage{graphicx}
\usepackage{dcolumn}
\usepackage{bm}
\usepackage{setspace}
\usepackage{amsmath}
\usepackage{comment}

\begin{document}


\title{Entropy maximization underlies topology and mechanical properties in dynamic covalent hydrogels}


\author{Lucien Cousin}
\affiliation{Macromolecular Engineering Lab, Department of Mechanical and Process Engineering, ETH Zürich, 8092 Zürich, Switzerland}
\author{Pietro Miotti}
\affiliation{Institute of Computing, Faculty of Informatics, USI Lugano, Switzerland}
\author{Bruno Marco-Dufort}
\affiliation{Macromolecular Engineering Lab, Department of Mechanical and Process Engineering, ETH Zürich, 8092 Zürich, Switzerland}
\author{Igor V. Pivkin}
\affiliation{Institute of Computing, Faculty of Informatics, USI Lugano, Switzerland}
\affiliation{(Swiss Institute of Bioinformatics, Geneva, Switzerland}
\author{Mark W. Tibbitt}
\affiliation{Macromolecular Engineering Lab, Department of Mechanical and Process Engineering, ETH Zürich, 8092 Zürich, Switzerland}


\date{\today}

\begin{abstract}


Adding dynamic bonds in polymer networks enables reprocessing and recycling; however the full impact of reversible bonds on dynamic network mechanics remains unclear. We build model dynamic networks and observe substantial deviations from classic theory. We rationalize these findings by considering that bond exchange enables the networks to rearrange and adopt a topology with a higher entropy. This allows us to accurately predict the gel point and elasticity of the dynamic networks. Further, we show by controlling bond exchange that network rearrangement can dramatically alter the mechanical properties, even without loss of bonds. 
\end{abstract}


\maketitle

\clearpage

Dynamic covalent networks have emerged as a promising class of soft materials with versatile applications \cite{kloxin2013covalent,winne2019dynamic,webber2022dynamic}. The introduction of dynamic covalent bonds allows for advanced mechanical properties, including reprocessability, as pioneered with vitrimers \cite{VanZee2020,Zou2017}, self-healing, or shape-shifting \cite{Zheng2021}. More recently, dynamic covalent networks have been engineered for biomedical applications, including for the thermal stabilization of biologics, biosensors, or three-dimensional printing \cite{cousin2025dynamic,BrunoBE,marco2022thermal}. In all of these applications, the key feature of dynamic covalent networks is their ability to combine the reversible nature of physical networks with the design possibilities and high mechanical strength of permanent covalent networks \cite{Zhang}. In this context, the precise understanding of classic properties of polymer networks, such as gel point and elasticity \cite{GelsreviewElastKey,SakaiBook}, are crucial in the engineering design of dynamic covalent networks. While an extensive body of literature has been devoted to the understanding of these quantities for the case of permanent covalent networks \cite{rent,SakaiBook}, less focus has been devoted to dynamic networks. 

Seminal work on the fundamental behavior of dynamic networks revealed the importance of bond thermodynamics and kinetics on the emergent properties \cite{Semenov1998,Rubinstein1998}. This was expanded upon to show that the gel point is determined by the fraction of bonds formed \cite{Sheridan2012}. Several frameworks have since been proposed to relate the elasticity and gel point of dynamic systems to their structure and the fraction of bonds formed \cite{Parada2018,BrunoM,craig2005}. 
%
These frameworks all rely on classic gelation theories, which assume that all reactive groups react independently from one another, and agree well with experimental data when most of the bonds are formed. However, dynamic networks often exhibit lower conversions---changing the proportion of bonds formed is often used to tune key mechanical properties, such as the elastic modulus, the ability to flow, to self-heal, or to respond to a stimulus \cite{kloxin2013covalent,cousin2025dynamic,webber2022dynamic}. In addition, the presence of unreacted bonds facilitates bond exchange, enhancing the dynamic behavior of dynamic covalent networks. In this context, dynamic networks are often applied below full conversion, but still above the gel point, motivating the need to better understand the physics in this regime \cite{Fujiyabu2019b,Parada2018}.

While previously proposed frameworks all rely on concepts originally developed for permanent networks \cite{Parada2018,BrunoM}, dynamic covalent networks possess the unique ability to rearrange even after gelation, enabling them to adopt internal structures that differ from that of their permanent counterparts. We hypothesized that this rearrangement allows dynamic networks to evolve toward thermodynamic equilibrium by adopting an internal structure associated with a higher entropy. 
We systematically varied the fraction of bonds formed, $p$, from $\approx 0.5$ to $\approx 1$. We measured the mechanical properties for each system with shear rheometry to determine the gel point and the evolution of the modulus as a function of $p$. We observed striking deviations in both properties from classical frameworks for polymer networks, in particular close to the gel point. By considering the entropy-maximized structure, we accurately predict the observed gel point and elasticity, enabling robust engineering of network properties across a broad range of $p$. We further confirm the importance of network structure by triggering bond exchange in an otherwise permanent network without changing conversion, and observe drastic changes in elasticity. This implies that reprocessable networks based on dynamic bonds may experience major changes in their mechanical properties after being reprocessed, even without loss of bonds.   

\begin{figure}
\includegraphics{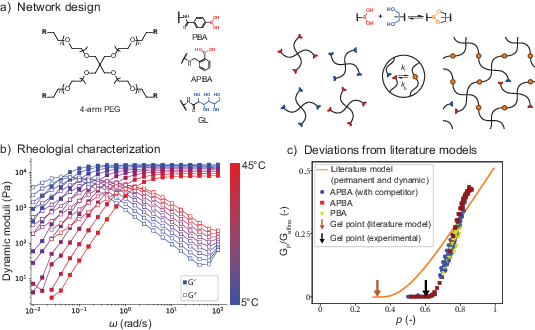}
\caption{\begin{setstretch}{1.083}
\textbf{Synthesis and mechanical characterization of dynamic covalent networks.}\label{fig:fig1} a) General structure of the dynamic covalent networks formed by cross-linking 4-arm PEG stars end-functionalized with either a boronic acid or a diol. The formed networks incorporate reacted and unreacted bonds in variable proportions, defined by $p$, the fraction of formed bonds. b) Chemical structure of the network precursors: 4-arm PEG stars end-functionalized either with phenylboronic acid (PBA) or aminophenylboronic acid (APBA) or with gluconolactone (GL) c) Frequency sweeps of networks formed with APBA and GL at 10 wt\% for temperatures ranging from 5$^{\circ}$C to 45 $^{\circ}$C. d) Normalized dynamic modulus $G'/G_{affine}$ as a function of $p$ for all conditions tested: varying temperature and concentration of network precursors for two different boronic acids, and varying concentration of a competitor small molecule and temperature for APBA. The orange line indicates the literature prediction \cite{Parada2018,BrunoM}.
\end{setstretch}}
\end{figure}

In this work, we formed model dynamic covalent networks (DCvNs) using 4-arm PEG stars, commonly called Tetra-PEG (Fig. \ref{fig:fig1}a). Covalent analogs of these networks have been shown to possess minimal defects compared with other known network architectures, making them suitable candidates to study the influence of reversible bonds independent of the properties of the network \cite{Matsunaga2009a,Nishi2014}. Dynamic bonds were introduced via the reversible binding of a boronic acid (BA) with a diol, in our case gluconolactone (GL) (Fig. \ref{fig:fig1})\cite{Brooks2018,BrunoBE}. We systematically varied $p$ from $\approx 0.5$ to $\approx 1$ by shifting the chemical equilibrium of the dynamic bonds. We quantifed the mechanics of each network using shear rheometry by performing frequency sweeps. 

As expected, the networks behaved like ideal Maxwell elements (Fig.\ref{fig:fig1}c)\cite{Semenov1998,Rubinstein1998}. DCvNs exhibited liquid-like behavior at low frequency, indicated by $\tan(\delta)<1$, and solid-like behavior at high frequency, with a cross-over frequency $\omega_c = \frac{1}{\tau_M}$, where $\tau_M$ is the characteristic relaxation time of the Maxwell material. The storage modulus increased as $\omega^{2}$ for low frequencies ($\omega < \omega_c$) and exhibited a plateau for high frequencies ($\omega > \omega_c$), consistent with scaling predictions\cite{Semenov1998,Rubinstein1998}. This agreement with scaling predictions indicated that the observed mechanics were only due to the dynamic bonds. This confirmed that the formed networks allowed us to isolate the impact of dynamic bonds from other network properties. We measured the storage modulus in the high frequency regime for $\omega \gg \omega_c$. In this regime, the dynamic bonds exchange at a rate much slower than the oscillation frequency and the storage modulus does not depend on $\omega$. Importantly, the elastic modulus in this regime is expected to be identical to that of a permanent network with the same internal structure \cite{Parada2018,Rubinstein1998}. 

In order to better understand the underlying features that govern the elastic modulus in our DCvNs, we first considered classic theories of rubber elasticity. In classic rubber elasticity, the elastic modulus of a polymer network arises from the sum of the elastic contributions of all polymer strands between cross-links \cite{rent,SakaiBook}. These theories primarily assume Gaussian chains whose elastic contribution is proportional to temperature. 
In order to take into account the role of conversion, some models consider in a first approximation that strands not connected to the infinite, percolated network do not contribute to the overall elasticity, while all connected strands have the same contribution \cite{Parada2018,Nishi2012}. In this approach, determining the elasticity of polymer networks is equivalent to counting the number of connected (also referred to as elastically active) strands. Based on these assumptions, the number of bonds formed, $p$, governs the elasticity of polymer networks. This approach has shown good agreement for ideal permanent networks \cite{Nishi2012} and dynamic covalent networks close to $p = 1$ \cite{Parada2018,BrunoM}. However, deviations have been shown for networks incorporating loops, or close to the gel point \cite{Lin2018,Akagi2013}, and more complex theories consider the influence of the local or global structure of the network \cite{rent,lange2011connectivity,BrassartPreStretch}. This structure is inferred from various considerations, such as the architecture and reactivity of the network precursors. However, the precise effect of different structures is still not fully understood, and was mostly investigated close to full conversion. We thus decided to investigate the role of $p$ in dynamic networks for $p$ ranging from at or below the expected gel point to near full conversion. 

In our system, we varied $p$ by systematically varying the following parameters: (i) the temperature of the gels; (ii) the concentration of the precursors (3 to 18 wt\%), keeping the stoichiometry constant at a value of 1:1; and (iii) the concentration of a competitive binder, keeping the concentration and stoichiometry of both precursors constant. To broaden the dataset, we formed networks using two different boronic acids (Phenylboronic acid, PBA, and 4-Aminophenylboronic acid, APBA). We measured the chemical equilibrium constants of each boronic acid and of the competitor with gluconolactone by Isothermal Titration Calorimetry (ITC). For each condition tested, we inferred the proportion of bonds formed, $p$, from the chemical equilibrium constant, and measured the plateau modulus $G_P$ and the mechanical relaxation time $\tau_M$ using shear rheometry. Experimental sets (i) and (ii) varied $p$ together with the elastic contribution of single chains and the total concentration of strands, respectively, while (iii) changed $p$ without influencing other parameters.

In order to isolate the effect of $p$ on the elasticity of dynamic networks, we normalized the measured elasticity by the prediction from the affine network model $G'_{affine}$, which does not include any effects due to deviations from an ideal network structure (Fig. \ref{fig:fig1}c). All measurements collapse on a single master curve, regardless of how the variation of $p$ was achieved. This suggested that $p$ was the main parameter governing the elastic modulus of DCvNs. Remarkably, this indicates that the deviations from the affine network model only depend on $p$ and not on how $p$ was varied, suggesting that $p$ may govern the internal structure of the DCvNs. In particular, temperature and concentration seem to influence the structure of DCvNs only through their influence on $p$. The collapse of the datasets on a single master curve suggests that networks with the same $p$ exhibit the same internal structure, even if they were made at different concentrations or temperatures.

We compared our experimental results with similar experiments performed on permanent networks formed using similar precursors and previously published models (Fig. \ref{fig:fig1}c). This comparison allowed for two observations: (i) the gel point was higher than predicted, and (ii) the increase in $G_P$ as a function of $p$ was steeper than predicted. The gel point is usually understood in the framework of the Flory--Stockmayer theory, which assumes that all reactive groups react independently of each other. For network precursors with functionality $f$ and 1:1 stoichiometry, it predicts $p_{gel} = \frac{1}{f-1}$  \cite{SakaiBook}. This gives $p_{gel} = \frac{1}{3}$ in our situation ($f=4$), while we observe $p_{gel} \approx 0.6$. We first rule out two explanations for this discrepancy. The plateau modulus was measured at frequencies high enough that the dynamic bonds were effectively frozen. We may thus rule out explanations based on the relaxation of the chains due to bond detachment. Some studies have suggested the role of chain pre-stretch in determining the elastic modulus of polymer networks, which may indicate that $G_P$ tending to $0$ is not an accurate measure of the gel point \cite{LangPreStretch,BrassartPreStretch}. To test this, we further measured $p_{gel}$ by interpolating the mechanical relaxation time $\tau_M$. We observed that $\tau_M$ scaled affinely with $p$ (see ESI). Following previous work, we identified $p_{gel}$ as the x-axis intercept \cite{Semenov1998,Sheridan2012}. We obtained the same value of $p_{gel} \approx 0.6$ as when analyzing $G_P$, corroborating our measurement of $p_{gel}$.

In order to investigate the physical origin of this delayed gelation, we considered the mechanism of gelation in DCvNs. In contrast to permanent networks, the stars and arms can rearrange after gelation, allowing them to explore different configurations. We hypothesized that the final configuration of the network (i.e., the configuration that is measured during rheometry) will converge to the configuration that maximizes the number of microstates. In other words, the system will maximize the entropy of the network given the constraints of the network. Given that chain pre-stretch was excluded, we did not consider the conformational entropy of the chains. We focused on the entropy associated with the connectivity of the stars that form the network. We stress that this connectivity does not depend on the physical properties of the chain. Instead, it can be represented in a mathematical or abstract way, for instance using a graph or a matrix. We refer to this quantity in the following discussion as the "network entropy". 

We first investigated whether network entropy could influence the chemical equilibrium of the bonds. Chemical equilibrium results from a competition between the energy of reaction and the spatial entropy of both reactants. Spatial entropy scales with the volume of the sample, while we expect network entropy to scale with the number of stars. Hence, network entropy should be much smaller than the spatial entropy of the reactants and should not affect chemical equilibrium. Based on this, we expect that network entropy will not affect $p$ directly. However, multiple configurations are possible for a given value of $p$, each associated with a different network entropy. Going from one of these configurations to another will leave $p$, and hence chemical equilibrium, unchanged, but will change the network structure and entropy. Chemical equilibrium and network entropy can thus be seen as two distinct and independent mechanisms governing network structure. Chemical equilibrium affects the proportion of bonds formed, $p$, without dictating the topology of the network, while network entropy affects the topology of the network for a fixed $p$. 

We then sought to quantify how the configuration of the network influences the entropy associated with the connectivity of the stars, i.e., how many microstates are accessible for a given configuration. For this, we need to identify relevant parameters to describe a configuration of the network. Previous works have considered that $p$ is enough to describe the network configuration in the case where all groups react independently \cite{nishi2017experimental,Parada2018}. In this case, the distribution of arms connected to the infinite network can be calculated using a framework developed by Miller and Macosko \cite{Miller1976}.
However, in DCvNs, the number of arms bound per star is not a fixed quantity and varies with time. Hence, we considered the probability distribution of the number of arms bound per star and write $P_i$ the probability for one star to have $i$ arms bound. This can be seen as a generalization of $p$, which can then be calculated as $p =\frac{1}{f}\sum_{i=0}^{f} iP_i$. We hypothesized that, among all possible probability distributions, the observed $P_i$ in dynamic networks should be associated with the highest entropy. This implies that in our DCvNs reactivity is not defined at the scale of a single reactive group, but at the scale of a star. In other words, reactive groups belonging to the same star are no longer considered independent. However, we considered that the stars are independent of each other: the probability distribution $P_i$ is the same for all stars, regardless of their environment. 

To calculate the number of microstates associated with a given distribution $P_i$, we recalled that these microstates do not depend, in our hypothesis, on the physical properties of the chains, but only on the connectivity of the stars. Hence, we considered a representation of the network as a graph. In this representation, only the information about the connectivity of the stars is kept, while the physical properties of the chains are disregarded. The vertices of the graph are the centers of the polymer stars and the edges are the polymer chains connecting two stars, i.e., the formed bonds. The number of accessible microstates can then be understood as the number of graphs that respect the constraints set by $P_i$. While this remains an open question in the field of graph theory, McKay and Wormald gave an asymptotic approximation, noted $G(\mathbf{k})$, of the number of graphs with a given degree sequence $\mathbf{k}$ for the case where the maximal degree is much lower than the total number of edges \cite{McKay1991}. This is the case here, as the maximal degree is 4 and the number of edges approaches the thermodynamic limit. The number of graphs in our case will then be $G(\mathbf{k})$ multiplied by the number of degree sequences that follow the probability distribution $P_i$. For a total number of stars $n_S$, a degree sequence follows the distribution $P_i$ if there are $n_S P_i$ occurrences of each degree in the sequence. The number of such sequences is given by the number of arrangements $\frac{n_S !}{\prod_{i}(n_S P_i)!}$. In addition, for each star of degree $k_i$, there are $\binom{4}{k_i}$ ways to choose which arms are connected. The full details of this derivation are given in the ESI. We then write the entropy as 
\begin{equation}
    \label{S1}
    S = k_B \log\left( G(\mathbf{k}) \frac{n_S!}{\prod_{i=0}^4 (n_S P_i)!} \prod_{i=0}^4 \binom{4}{i}^{n_S P_i} \right)
\end{equation}

We hypothesized that the probability distribution $P_i$ in dynamic networks should maximize this entropy, given the constraint that $p$ is already known. We calculated this probability distribution using Lagrange multipliers (see ESI).

In order to derive the gel point, we need to calculate the number of arms bound to the infinite network. We used the recursive approach developed by Miller and Macosko \cite{Miller1976}. Importantly, their derivation relies on the key assumption that all reactive groups are independent of each other and react with equal probability $p$. In the case of dynamic interactions between stars, we posit that the stars---and not the arms---are independent of each other and adapt the derivation accordingly. Following the notation of Miller and Macosko, we consider the probability $P(F^{\text{in}})$ that when entering a star through a connected arm, none of the arms lead to the infinite network. In this notation, $F$ refers to the inclusion in a finite chain, while the superscript $in$ indicates that the probability is taken when entering an arm. $P(F^{\text{in}})$ can be calculated recursively by the probability that when entering all neighboring stars, all of their arms do not lead to the infinite network. The equation for $P_{in}$ follows: 
\begin{equation}
    P(F^{\text{in}}) = \sum_{i=1}^4 \dfrac{P_i}{\sum_{i=1}^4 P_i}P(F^{\text{in}})^{i-1}
\end{equation}
 where we need to pay attention to the fact that the probability for a star to have $i$ connections is  $\frac{P_i}{\sum_{i=1}^4 P_i}$ because we know that there is at least one arm bound. From there, we write the probability that $n$ arms are connected as: 
\begin{equation}
    \mathrm{P}(n\text{ arms connected}) = \sum_{i=n}^4P_i\binom{i}{n}P(F^{\text{in}})^{i-n}\left(1-P(F^{\text{in}})\right)^n
\end{equation}
While this recursive approach follows a similar logic to that developed by Miller and Macosko, the key assumption that reactivity can be modeled at the scale of a single reactive group breaks down in our case. Maximization of network entropy requires in our approach that the probability distribution $P_i$ is determined by eq. \ref{S1}, which only considers the number of bonds per star. Because entropy maximization occurs at the scale of the network as a whole, this assumption is likely to be a first-order approximation, and one could imagine that structures on the scale of multiple stars also determine the number of microstates.

\begin{figure}
\includegraphics{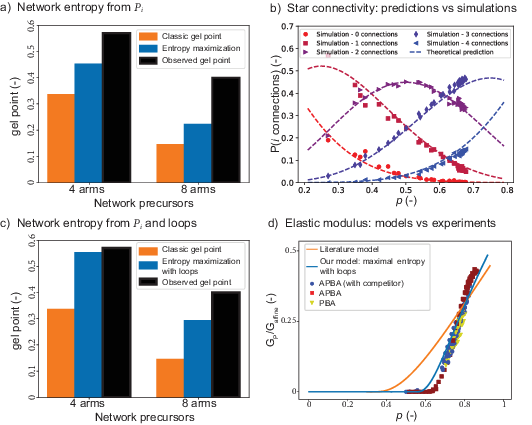}
\caption{\label{fig:fig2} \begin{setstretch}{1.083}
\textbf{The maximization of network entropy in dynamic covalent networks leads to a different connectivity that explains the measured gel point.} a) Measured gel point in the case of 4-arm and 8-arm stars compared with the predictions from the Flory--Stockmayer model and with our prediction, taking into account network entropy maximization calculated from $P_i$. b) Calculated fraction of stars with $i$ arms bound, $i = 0 \cdots 4$ (dashed lines), compared to the simulated fraction (symbols). c) Measured gel point in the case of 4-arm and 8-arm stars compared with the predictions from the Flory--Stockmayer model and with our prediction, taking into account network entropy maximization calculated from $P_i$ and from the fraction of loops. d) Normalized dynamic modulus $G'/G_{affine}$ as a function of $p$ for all conditions tested (full symbols) compared with the literature model (orange line, \cite{Parada2018,BrunoM}) and with our model (blue line).
\end{setstretch}}
\end{figure}

The gel point predicted using this connectivity resulting from entropy maximization is substantially increased ($p_{gel} = 0.45$) compared to previous approaches ($p_{gel} = 0.33$) (Fig. \ref{fig:fig2}a). However, it still does not capture the experimental results. This suggests that other factors beyond $P_i$ may contribute to network entropy. The recursive approach used to calculate the gel point assumes an ideal structure in which none of the possible paths exiting a star will lead back to it. As the dynamic bonding provides freedom for the system to rearrange during entropy maximization, we hypothesized that this process could result in deviations from ideal network structure. For example, allowing paths to loop back to the star they originate from could lead to a greater number of microstates, and hence increased network entropy. These deviations from an ideal network structure are often called defects, and substantially impact the overall elasticity \cite{lange2011connectivity,rent}. We modified our approach by accounting for loops between adjacent stars, which are expected to have the greatest impact on the overall elasticity \cite{rent}. We define 2-loops when two stars have 2 arms bound together, and more generally i-loops when two stars have $i$ arms bound together. Networks formed from two types of 4-arm stars have been shown to have a nearly ideal structure and various attempts have been made to quantify the number of defects in such networks \cite{lange2011connectivity,olsenNDS,lin2018topological}. However, the proportion of loops found either using advanced experimental tools or via numerical simulations in these covalent networks cannot account for the delayed gelation in our case, as they still exhibited a gel point $p_{gel} = 0.33$ \cite{nishi2017experimental}.

As loops also contribute to network entropy, we hypothesized that they may be present in different proportions in dynamic and permanent networks. We further hypothesized that the proportion of loops is determined by the maximization of network entropy. Hence, we calculated the proportion of $i$-loops using the same approach as we used to calculate the distribution $P_i$: first, we derived a relation linking the proportions of $i$-loops and network entropy; then, we calculated the proportions of $i$-loops that maximize network entropy. 

In order to calculate the entropy associated with loops, we built upon our approach developed above based on McKay and Wormald's theorem. This result does not consider loops. Thus, we assumed that the distribution $P_i$ is set by the maximization of entropy according to Eq. \ref{S1}, and considered a second abstract representation of the network. We consider bonds as individual items that can form pairs. This possibility to form pairs will result in a different number of possible microstates depending on the number of pairs formed. In this representation, the bonds are now assumed to be indistinguishable: this is necessary to avoid counting the same microstates in the enumeration of graphs and in the counting of loops. This results in a different, higher-level representation of the network than was used previously: instead of resolving the network star by star, we consider parameters at a global scale, such as the probability distribution associated with the connections of the stars and the number of loops in the network. In our approach, both representations offer complementary descriptions of the network structure and the associated microstates. We provide here the idea of the derivation of the proportion of loops for the case where only 2-loops are considered. The full details can be found in the ESI.

We write $n_l$ the number of 2-loops and $n_b$ the number of bonds. The number of possible states is then the number of ways to choose the $2n_l$ bonds that will form 2-loops, multiplied by the number of ways to arrange these bonds into $n_l$ 2-loops. The first term can be quantified by the combinatorial coefficient $\binom{n_b}{2n_l}$. The second term can be understood as the number of matchings in a complete graph of $2n_l$ vertices, and is given by $\frac{(2n_l)!}{2^{n_l}n_l!}$. In addition, we must consider that the loops themselves are indistinguishable, and divide the result by $n_l!$. This last term can also be understood as the number of degrees of freedom lost when pairing bonds: because the system will remain the same under permutation of all loops, fewer distinct microstates are available. In this view, the equilibrium configuration of the system can be seen as the result of a competition between the microstates created by pairing bonds and the microstates deleted by constraining some bonds to be paired with another. We then write the entropy as 
\begin{equation}
    S = k_B \left( \log \binom{n_b}{2n_l} +\log\left( \dfrac{(2n_l)!}{2^{n_l}n_l!}\right) -\log (n_l!) \right)
\end{equation}
 We then consider that the proportion of loops is the value of $\frac{n_l}{n_b}$ that maximizes $S$ in this equation. We use a very similar approach to calculate the proportions of 3-loops and 4-loops, and the derivation is provided in the ESI. The proportions are plotted in Fig. \ref{fig:fig2}. 

To corroborate our theoretical prediction, we also performed numerical simulations of DCvNs using dissipative particle dynamics \cite{pietroDPD}. The distribution $P_i$ calculated with our approach agrees well with the results obtained from numerical simulations (Fig \ref{fig:fig2}b). In addition, we compared the probability that one arm is not connected to the infinite network with the results from numerical simulations and observed good agreement. Thus, the theoretical and computational analyses support the hypothesis that DCvNs exhibit a different internal structure than their permanent counterparts, which further motivates our approach to explain the changes in gel point and modulus evolution. 

Knowing the distribution $P_i$ and the proportion of $i$-loops, we determined the gel point using the same recursive approach as before. We include the effect of loops by considering that a loop is equivalent to a single link between two stars: for instance, a star with 3 arms bound including one 2-loop is topologically equivalent to a star with 2 arms bound. We modify the distribution $P_i$ accordingly using the proportions of $i$-loops and use the recursive approach described above. The gel point predicted considering the topology resulting from entropy maximization with loops agreed well with the experimental data (Fig. \ref{fig:fig2}c), further corroborating our hypothesis that the different internal structure due to the increase in network entropy drives changes in the mechanical properties of dynamic networks. 

Next, we sought to investigate if these changes in network structure could also explain the steeper increase in $G'$ after gelation in DCvNs. As our previous analysis showed that both are required to explain the shift in the gel point, a model for the elasticity of DCvNs should include the effects of dangling chains and loops. In addition, it would have to be applicable close to gelation, where the proportion of dangling chains is expected to be high. Some models have been published to explain the influence of several types of defects or the proximity to the gel point on the elasticity of polymer networks \cite{LangPreStretch,rent,nishi2017experimental}. However, it remains unclear how to include both the proximity to the gel point and the presence of loops and dangling chains in a single model.  

We thus developed a simple scaling model for the elasticity of polymer networks based on the Phantom Network Model \cite{SakaiBook}. This model assumes that the network is tree-like and that each polymer segment is connected to the boundary of the network via an infinite tree. For Gaussian chains, the elasticity of this infinite tree can be modeled as the elasticity of a chain of length $N_\infty$, which can be derived from the length of the chains between the cross-links. Following the reasoning developed in RENT to include the effect of various types of defects in the calculation of $N_\infty$ \cite{rent,rentPreStrain}, we propose to include the effect of loops and dangling chains via a mean-field approach. More precisely, we consider that network defects change the effective length of the strands between cross-links as compared with an ideal network. We calculate the average length between two cross-links as a function of the proportions of dangling chains and loops. We then derive the equivalent length $N_\infty$ and the elastic modulus based on the calculations of the Phantom Network Model. While previous approaches focused on the functionality of the cross-links, our model includes both the functionality of the cross-links and the effective length of the strands to link the proportion of defects with the macroscopic elasticity. 
The details of the derivation can be found in the ESI. 

Our prediction for the elasticity of DCvNs as a function of $p$ that accounts for internal structural changes due to network entropy maximization and the associated presence of loops (blue line) predicts both the delayed gel point and the increase in $G_P$ with $p$ (Fig. \ref{fig:fig2}d). Notably, established models in the literature for dynamic networks (orange line) that build upon classic network theory, which account for dangling chains but not loops or network entropy maximization \cite{Parada2018}, predict a lower gel point than observed and a less steep increase in modulus with increasing $p$. 
The ability of our model to capture the gel point and relation between modulus and extent of reaction indicates that accounting for network entropy maximization accurately captures the structure of the network, both in terms of the connectivity of the stars and of the defects in the network. As a consequence, this corroborates our hypothesis that the arms of a star are not independent in dynamic polymer networks, which is a key difference from our understanding of permanent polymer networks. Some deviations between our model and the experimental data remain for $p \gtrsim p_{gel}$ and $p \lesssim 1$, which may be attributed to additional structural changes that were not taken into account in our calculations, for instance loops involving more than two stars. These are expected to both further delay the gel point and stiffen the network once formed. However, our framework could account for and model the influence of these, or other types of structures. Similar to loops, their proportion should be dictated by the maximization of network entropy. The fraction of each type of defect could thus be calculated using the same framework as before: expressing network entropy as a function of the fraction of each type of defect, then finding the fraction of defects that maximizes this expression. The effect on long-range connectivity could then be included in our recursive approach. Such calculations would however be highly complex and could be the subject of future work.

These results highlight that the knowledge of $p$ and the topology of the precursors alone are not sufficient to predict the mechanical behavior of dynamic polymer networks. This holds even when the system is probed at frequencies above $1/\tau_R$ where the individual bonds are effectively frozen. In contrast, this information ($p$ and precursor topology) is generally sufficient to describe the properties of ideal permanent covalent networks, indicating that DCvNs exhibit important differences in their underlying structure and connectivity as compared with their permanent counterparts for the same $p$. This is most strikingly seen for networks with $p \lesssim p_{gel} \approx 0.6$---while permanent networks are past the gel point and exhibit an elastic modulus in the order of a few kPa, DCvNs are below or close to their gel point and fail to form a network or exhibit nascent gel formation with an elastic modulus of a few Pa.

\begin{figure*}
\includegraphics{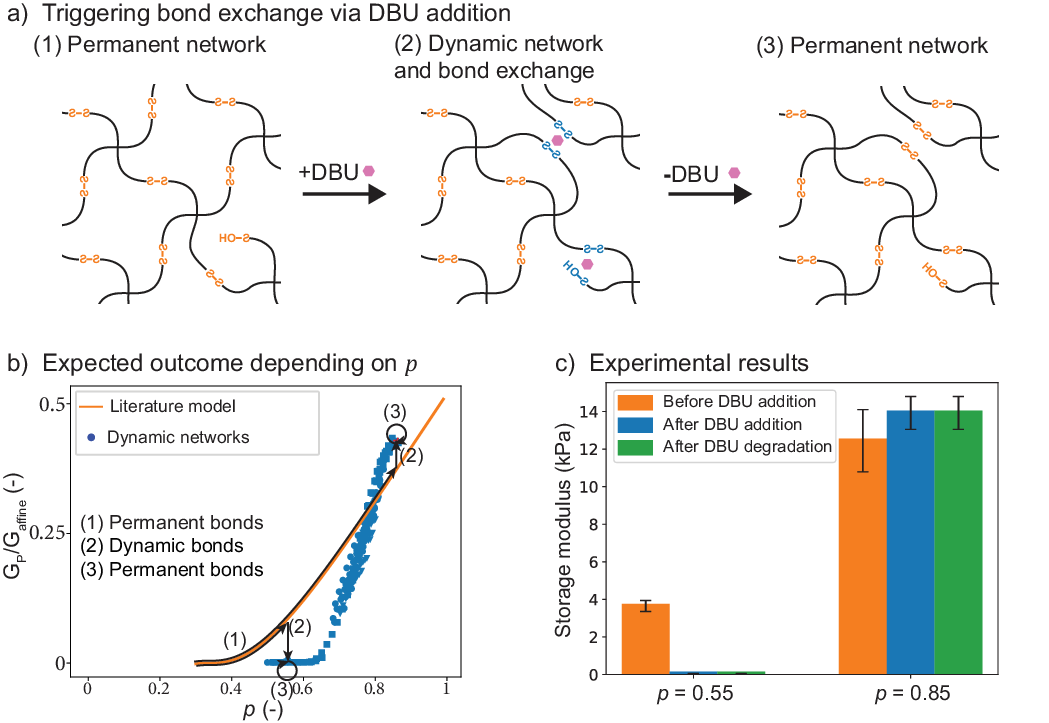}
\caption{\label{fig:fig3}  \begin{setstretch}{1.083}
\textbf{Triggering bond exchange leads to a persistent change in the mechanical properties of transiently reversible networks.} a) Schematics of the experiment: starting with (1) permanent disulfide bonds, DBU is added (2) to trigger the exchange of the disulfide bonds (highlighted in blue). Upon degradation of DBU (3), the disulfide bonds become permanent again. Unreactive OH groups create dangling ends. b) Expected outcome based on our results and literature models depending on $p$. c) Average and standard deviation of the storage moduli for networks with $p=0.55$ and $p=0.85$ during phases (1), (2), and (3).
\end{setstretch}
}
\end{figure*}

To validate these theoretical concepts experimentally and show that $p$ is not enough to understand the mechanical properties of polymer networks, we created polymer networks with the same $p$ but different elastic moduli. We achieved this by creating polymer networks with triggerable bond exchange, allowing us to enable bond exchange in effectively permanent networks and turn it off so that the networks behave effectively permanent again (Fig. \ref{fig:fig3}). If our hypothesis is correct, triggering bond exchange should lead to a change in network structure as a result of network entropy maximization, and to a change in $G'$. Further, this change should remain when the bonds are made effectively permanent again, as it is only an effect of the structure. For this, we created networks based on 4-arm PEG stars linked together by disulfide bridges and varied $p$ by introducing varying equivalents of non-reactive -OH groups that form dangling chains (Fig. \ref{fig:fig3}). The disulfide bridges were effectively permanent at room temperature and neutral pH, but were made dynamic by adding the base DBU \cite{DBUSS,DBUSS2}. Degradation of DBU over time made the bonds effectively permanent again. The use of DBU to make disulfide bonds dynamic has been reported previously in the literature \cite{DBUSS,DBUSS2} and was assessed by oscillatory rheology in our system (see ESI).

Based on this strategy, we formed networks with $p=0.55$ and $p=0.85$. Based on theoretical predictions \cite{nishi2017experimental}, we would expect that the effectively permanent network with $p=0.55$ would form with a modulus of 4 kPa and upon triggering dynamicity via the addition of DBU the network modulus should decrease substantially or the network should dissolve. Further, these properties should remain after the degradation of DBU when the bonds revert to an effectively permanent case. In contrast, the network with $p=0.85$ should stiffen upon triggering dynamicity. The measured $G'$ were consistent with our hypothesis (Fig. \ref{fig:fig3}c). Upon addition of the base DBU, the storage moduli of the materials decreased ($p=0.55$) or increased ($p=0.85$) to match the moduli that we observed for DCvNs with the same $p$. That is, polymer networks with triggered dynamicity prepared with $p = 0.55$ liquefied and those prepared with $p = 0.85$ stiffened. The moduli remained the same after DBU degraded and the networks became permanent again, highlighting that the changes to the networks remained. This further supports the model that dynamic bonding in polymer networks gives access to internal structures that are substantially different from permanent networks. In addition, this shows the crucial role of network structure in determining the mechanical behavior of polymer networks: for $p=0.55$, a change in structure alone leads to the liquefaction of the network. In order to observe the same change by varying $p$ without affecting the structure, one would have to decrease $p$ by more than 0.2, i.e, break more than a third of the formed bonds in the material. The fact that $G'$ changes in opposite directions depending on $p$ suggests that bonds are neither broken nor created during this process and that the observed changes are a result of the internal structure, as we predicted. 

We highlight that the effect evidenced by these experiments is purely due to the connectivity of the network precursors and should not depend on the chemistry used to form the bonds, nor on the stimulus used to make the bonds dynamic. This implies that other types of networks could experience the same effect. Covalent Adaptable Networks (CAN) have emerged as a promising class of value-added materials, in which the covalent bonds linking the network precursors become dynamic under the application of a stimulus. Our observations indicate that such networks could experience changes, in some cases major changes, in their mechanical properties when the covalent bonds are made dynamic. For instance, a CAN prepared in the permanent state with $p=0.55$ will turn into a liquid when the bonds are first made dynamic, while $p$ remains unchanged. 

In this letter, we showed that the ability of DCvNs to rearrange after gelation allows them to adopt internal structures that maximize network entropy. Network entropy is here quantified as the information entropy of the network and does not depend on the conformational entropy of the individual chains. In the case of the networks investigated here, this leads to a substantially delayed gel point and a steeper increase in $G'$ after gelation. We explained the delayed gel point by considering the increased proportion of loops and by showing that the network precursors exhibit a different connectivity at a larger scale. This, combined with a modified model for rubber elasticity, allowed us to reproduce the steeper increase in $G'$ after gelation. Our results demonstrate that, for a given $p$ and precursor architecture, different internal structures are possible, which may lead to widely different mechanical properties. This highlights that a careful determination of the network structure is essential to understand the macroscopic properties of polymer networks. This is particularly relevant for CANs, which may exhibit major and irreversible changes in their mechanical properties after the bonds are made dynamic. This aspect of network rearrangement and impact on macroscopic properties should be considered when engineering reprocessable or recyclable materials based on dynamic bonds. Depending on the system and conversion $p$, robust materials may not reform or may strengthen after rearrangement. In addition, this work introduces a new concept related to network entropy to the characterization of polymer networks and provides the physical and mathematical framework to apply it. We are just beginning to understand the impact of network entropy on network structure and properties, and these effects may have broader impact in the network community beyond the dynamic covalent networks explored here.
\section{Supplementary Information}


\section{Derivation of the distribution of the number of connected arms per star}

We start by giving an expression for the entropy associated with the topology of the network. In the following, we call this quantity "network entropy". As stated in the main text, we expect that this entropy is much smaller than the spatial entropy associated with the motion of the functional groups. Since chemical equilibrium results from a competition between this spatial entropy and the energy of binding, taking into account the entropy associated with the topology of the network should not change the chemical equilibrium significantly. Hence, we assume that the change in network entropy occurs at fixed $p$. 

We considered that network entropy originates from the number of microstates accessible for a given $p$. To describe the number of microstates, we represented the configuration of the network by a graph (Fig.[ref]): the vertices of the graph are the centers of the polymer stars, and the edges are the formed bonds between the polymer stars. In this representation, an edge links two vertices if and only if the two stars associated with the vertices are linked by a bond. Dangling chains are thus not represented, so the degree of each vortex varies between 0 and 4. We parametrize the network by using the probability distribution of the number of arms bound per star, referred to as $P_i$. Based on this representation, we considered that the number of microstates is the number of graphs that respect two conditions: 
\begin{itemize}
    \item The total number of bonds formed in the system is $n_S\times 4p$. 
    \item The total number of stars with $i$ arms bound corresponds to the distribution $P_i$
\end{itemize}

\begin{figure}
\includegraphics[width=\textwidth]{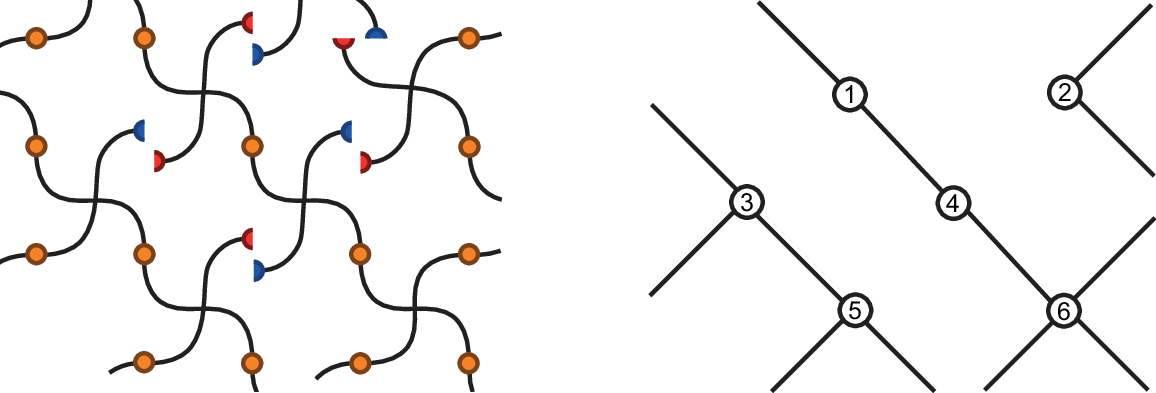}
\caption{\label{figfiggraph} Schematics of the graph representation used to calculate the number of accessible microstates. Left: schematic of a real network. Right: Graph representation associated with this network. }
\end{figure}

Quantifying the number of graphs that respect given constraints remains an open question in the field of graph theory, but McKay and Wormald have given an approximation for the case where the number of vertices is much greater than the maximal degree. This condition is clearly satisfied in our case, as the maximum degree is 4 and the number of vertices is equal to the number of stars in a network, which approaches the thermodynamic limit. 

The formula derived by McKay and Wormald gives an approximation of the number of graphs when the degree sequence is known, i.e., when the vertices $1,\cdots, n$ that compose the graph have degrees $k_1, \cdots, k_n$. Physically, this corresponds to associating each star with a number and writing $k_i$ the number of arms bound for the star $i$. Let $\mathbf{k}$ be the vector describing the degree sequence: $\mathbf{k} = \left( k_1, \cdots, k_n \right)$. The number of graphs that respect the degree sequence $\mathbf{k}$ is 
\begin{equation}
    G(\boldsymbol{k}) \sim \sqrt{2}\left(\lambda^{\lambda}(1-\lambda)^{1-\lambda}\right)^{\binom{n}{2}} \prod_{i=1}^{n}\binom{n-1}{k_{i}} \exp \left(\frac{1}{4}-\frac{\gamma_{2}^{2}}{4 \lambda^{2}(1-\lambda)^{2}}\right)
    \label{eqWM}
\end{equation}
In this expression, the following notations are used: 
\begin{equation}
\def\arraystretch{2.2}
\begin{split}
    \lambda &= \dfrac{1}{n_S-1} \bar{k} \\
    \gamma_2 &= n_S \bar{k}^2 \dfrac{\nu_2-1}{\left(n_S-1\right)^2} \\
    \bar{k} &= \dfrac{1}{n_S} \sum_{i=1}^{n_S} k_i\\
    \nu_2 &= \dfrac{1}{\bar{k}^2n_S} \sum_{i=1}^{n_S}k_i^2
\end{split}
\label{notationsWM}
\end{equation}
We now have to find the number of graphs that respect the two conditions listed above. With $n_S$ the number of stars in the system and $P_i$ the probability for a star to have $i$ arms bound, the number of stars that have $i$ arms bound is $n_S P_i$. This implies that each sequence $\mathbf{k}$ should be composed of $n_S P_i$ occurrences of $i$. In other words, the number of indices $j$ such that $k_j = i$ should be equal to $n_S P_i$ for $i \in [0,4]$. Let us write $n_i = n_SP_i$ in order to make the following equations more clear. The number of sequences $\mathbf{k}$ that respect this condition is given by the number of arrangements: 
\begin{equation}
    \dfrac{n_S!}{\prod_{i=0}^4 \left( n_i \right)!}
    \label{eqArrangements}
\end{equation}
In addition, we have to consider that for a given degree $k_i$, there are $\binom{4}{k_i}$ configurations possible for a star, corresponding to which arms are bound. This implies that for a given degree sequence $\mathbf{k}$, the number of configurations of the stars is
\begin{equation}
    \prod_{i=0}^4 \binom{4}{i}^{n_i}
    \label{eqArrangements2}
\end{equation}
We can now write the number of accessible microstates as the product of the number of graphs for a given degree sequence $\mathbf{k}$ with the number of configurations of the stars that lead to a degree sequence $\mathbf{k}$. Taking the logarithm to write the entropy, one arrives at:
\begin{equation}
    S = k_B \log\left( G(\mathbf{k}) \dfrac{n_S!}{\prod_{i=0}^4 (n_i)!} \prod_{i=0}^4 \binom{4}{i}^{n_i} \right)
    \label{SWM1}
\end{equation}
We now turn our attention to the function $\log G(\mathbf{k})$. In order to make it usable, we simplify it by considering that $n_S \gg k_i$ and express it as a function of $P_i$ instead of $k_i$.

We expand the logarithm in eq. \ref{SWM1} to get:
\begin{equation}
\begin{split}
\dfrac{S}{k_B} = & \log \sqrt{2} + \lambda \binom{n_S}{2}\log \lambda + (1-\lambda)\binom{n_S}{2} \log (1-\lambda) + \sum_{i=1}^{n_S}\log \binom{n_S}{k_i} + \dfrac{1}{4}-\dfrac{\gamma_{2}^{2}}{4 \lambda^{2}(1-\lambda)^{2}} \\ 
 & + \log \left( \dfrac{n_S!}{\prod_{i=0}^4 (n_i)!} \prod_{i=0}^4 \binom{4}{i}^{n_i} \right)
\end{split}
    \label{eqSWMlog}
\end{equation}
From eq. \ref{notationsWM} it is clear that $\lambda \ll 1$. we can rewrite the first term as:
\begin{equation}
\begin{split}
   \lambda \binom{n_S}{2}\log \lambda & = \dfrac{1}{2}\sum_{i=1}^{n_S} k_i \times \log \left( \dfrac{\sum_{i=1}^{n_S} k_i}{n_S (n_S-1)} \right) \\
    & \approx \dfrac{1}{2}\sum_{i=1}^{n_S} k_i \times \log \left( \sum_{i=1}^{n_S} k_i\right) - \sum_{i=1}^{n_S} k_i \times \log \left( n_S  \right)
\end{split}   
\end{equation}
Similarly, we can rewrite the second term as 
\begin{equation}
\begin{split}
   (1-\lambda)\binom{n_S}{2} \log (1-\lambda) & \approx (1-\lambda)\binom{n_S}{2} (-\lambda) \\
    & \approx -\binom{n_S}{2} \lambda \\
     & = -\dfrac{1}{2}\sum_{i=1}^{n_S} k_i
\end{split}   
\end{equation}
This shows that these two terms are on the order of $\sum_{i=1}^{n_S} k_i$, i.e., on the order of $n_S$.
 
We now show that we can neglect the term $\dfrac{1}{4}-\dfrac{\gamma_{2}^{2}}{4 \lambda^{2}(1-\lambda)^{2}}$. For this, we first develop $\gamma_2$: 
\begin{equation}
    \begin{split}
        \gamma_2 & = n_S \bar{k}^2 \dfrac{\dfrac{\sum_{i=1}^{n_S} k_i^2}{n_S \bar{k}^2}-1}{(n_S-1)^2} \\
         & \approx \dfrac{1}{n_S^2} \left( \sum_{i=1}^{n_S} k_i^2 - n_S \bar{k}^2 \right) \\
          & = \dfrac{1}{n_S} \left( \overline{k^2}- \overline{k}^2 \right)
    \end{split}
\end{equation}
From this, we can simplify the following term: 
\begin{equation}
    \begin{split}
        \dfrac{\gamma_{2}^{2}}{ \lambda^{2}(1-\lambda)^{2}} & \approx \dfrac{\gamma_{2}^{2}}{ \lambda^{2}} \\
         & \approx \dfrac{1}{n_S^2} \left( \overline{k^2}- \overline{k}^2 \right)^2 \times \dfrac{1}{\dfrac{\bar{k}^2}{n_S^2}}\\
          & = \dfrac{\left( \overline{k^2}- \overline{k}^2 \right)^2}{\bar{k}^2}   
    \end{split}
\end{equation}
It then becomes clear that the term $\dfrac{1}{4}-\dfrac{\gamma_{2}^{2}}{4 \lambda^{2}(1-\lambda)^{2}}$ is on the order of $1 \ll n_S$. it can this be neglected compared to $\lambda \binom{n_S}{2}\log \lambda + (1-\lambda)\binom{n_S}{2} \log (1-\lambda)$.

We now turn our attention to the term $\sum_{i=1}^{n_S}\log \binom{n_S}{k_i}$, and simplify it using Stirling's approximation for the term $n_S!$:
\begin{equation*}
    \log \dfrac{n_S!}{k_i! \left( n_S-k_i\right)!} \approx n_S \log n_S-\left( n_S-k_i\right) \log \left( n_S-k_i\right)-\log (k_i!)-k_i
\end{equation*}
where we used Stirling's approximation for $\log n_S!$ and $\log \left( n_S-k_i\right)!$. We the use a first-order Taylor expansion, using the assumption $k_i \ll n_S$: 
\begin{equation*}
     n_S \log n_S-\left( n_S-k_i\right) \log \left( n_S-k_i\right) \approx k_i \log n_S + k_i
\end{equation*}
This allows us to rewrite the sum as 
\begin{equation}
    \sum_{i=1}^{n_S}\log \binom{n_S}{k_i} \approx \sum_{i=1}^{n_S} k_i\log n_S - \log k_i !
\end{equation}

Putting it all together, we can rewrite $\log G(\mathbf{k})$ as
\begin{equation}
     \log G(\mathbf{k}) \approx \dfrac{1}{2}\sum_{i=1}^{n_S} k_i \times \log \left( \sum_{i=1}^{n_S} k_i\right) - \sum_{i=1}^{n_S} k_i \times \log \left( n_S  \right) -\dfrac{1}{2}\sum_{i=1}^{n_S} k_i + \sum_{i=1}^{n_S} k_i\log n_S - \log k_i !
    \label{eqWMlogSimp}
\end{equation}

Taking the logarithm of eqs. \ref{eqArrangements} and \ref{eqArrangements2} is straightforward using Stirling's approximation and yields: 
\begin{equation}
\begin{split}
     & \log \left( \dfrac{n_S!}{\prod_{i=0}^4 (n_i)!} \prod_{i=0}^4 \binom{4}{i}^{n_i} \right) \\
      & \approx n_S \log n_S - \sum_{i=0}^4 n_i \log (n_i) + \sum_{i=0}^4 n_i \log \binom{4}{i}
\end{split}
\end{equation}
We can now combine the two previous equations to arrive at:
\begin{equation}
\begin{split}
    \dfrac{S}{k_B} \approx & \dfrac{1}{2}\sum_{i=1}^{n_S} k_i \times \log \left( \sum_{i=1}^{n_S} k_i\right) - \sum_{i=1}^{n_S} k_i \times \log \left( n_S  \right) -\dfrac{1}{2}\sum_{i=1}^{n_S} k_i + \sum_{i=1}^{n_S} k_i\log n_S - \log k_i ! \\
     & + n_S \log n_S - \sum_{i=0}^4 n_i \log (n_i) + \sum_{i=0}^4 n_i \log \binom{4}{i}
\end{split}
\end{equation}

We now notice that we can rewrite $\sum_{i=0}^{n_S} k_i$ as a function of $n_i$: 
\begin{equation*}
    \sum_{j=0}^{n_S} k_j = \sum_{i=0}^{4}i \times n_i
\end{equation*}
This gives an expression of $S$ as a function of the distribution $n_i$ alone: 
\begin{equation}
\begin{split}
    \dfrac{S}{k_B} \approx & \dfrac{1}{2}\sum_{i=0}^{4} in_i \times \log \left( \sum_{i=0}^{4} in_i\right) - \sum_{i=0}^{4} in_i \times \log \left( n_S  \right) -\dfrac{1}{2}\sum_{i=0}^{4} in_i + \sum_{i=0}^{4} in_i\log n_S \\
     &- \sum_{i=0}^{4}n_i\log i !  + n_S \log n_S - \sum_{i=0}^4 n_i \log (n_i) + \sum_{i=0}^4 n_i \log \binom{4}{i}
\end{split}
\label{eqSWMFin}
\end{equation}

We can now find the values of $n_i$ that maximize the entropy defined in eq. \ref{eqSWMFin}. Although our model focuses on the probability distribution $P_i$, finding the absolute values of the number of stars with $i$ arms bound, $n_i$, is equivalent and easier mathematically in this case. We will show at the end of the calculation that a simple normalization is enough to recover $P_i$.

The distribution $n_i$ has to satisfy two constraints. First, the total number of stars$n_S$ is known, giving $\sum_{i=0}^4 n_i = n_S$. Second, the total number of arms bound is known and is equal to $4pn_S$: there are 4 arms per star and each arm has a probability $p$ to be bound. This gives $\sum_{i=0}^4 in_i = 4p$. We use Lagrange multipliers, a mathematical method to find the maximum of a multivariate expression under constraints. In order to simplify the following expressions, we try to find the location of the maximum of $\frac{S}{k_B}$.

We first give the expression for the Lagrange function: 
\begin{equation}
    L = \dfrac{S}{k_B} + \lambda \left( \sum_{i=0}^4 P_i - 1 \right) + \mu \left( \sum_{i=0}^4 iP_i - 4p \right)
\end{equation}
Here, $\lambda$ and $\mu$ are constants and their value will be determined later, following the procedure of Lagrange multipliers.
The maximum of $S$ under the constraints defined above corresponds to an extremum of $L$. We can find the location of this extremum by equating all partial derivatives to $0$:
\begin{equation}
\def\arraystretch{1.7}
    \left\{
    \begin{array}{ll}
        \dfrac{\partial L}{\partial n_i} = 0 & \quad \text{ for }0\leq i\leq 4 \\
        \dfrac{\partial L}{\partial \lambda }= 0  &\quad \text{} \\
        \dfrac{\partial L}{\partial \mu }= 0 & \quad\text{} 
    \end{array}
    \right.
\end{equation}

The last two terms give the constraints on $P_i$. The first term can be rewritten as 
\begin{equation*}
    \dfrac{\partial L}{\partial n_i} = \dfrac{1}{2}i \left( 1+\log \sum_{i=0}^4 in_i \right)-i-\log i! -1-\log n_i +\log \binom{4}{i} +\lambda+\mu i = 0
\end{equation*}
This gives an expression of $\log n_i$:
\begin{equation*}
    \log n_i = i \left(\dfrac{1}{2}i \left( 1+\log \sum_{i=0}^4 in_i \right)-1+\mu \right) -\log i! +\log \binom{4}{i} -1+\lambda
\end{equation*}
Taking the exponential: 
\begin{equation}
    n_i = \dfrac{A^iB}{i!}\binom{4}{i}
    \label{niGen}
\end{equation}
Where $A$ and $B$ are functions of $\lambda$, $\mu$, and $p$. In order to determine their value and fully characterize $n_i$, we now solve the two equations corresponding to the two constraints: 
\begin{equation}
\def\arraystretch{1.7}
    \left\{
    \begin{array}{ll}
        \sum_{i=0}^4 n_i = \sum_{i=0}^4  \dfrac{A^iB}{i!}\binom{4}{i}= n_S & \quad (1) \\
        \sum_{i=0}^4i n_i = \sum_{i=0}^4  i\dfrac{A^iB}{i!}\binom{4}{i}= 4pn_S & \quad (2)
    \end{array}
    \right.
    \label{condLagrange}
\end{equation}
From condition $(1)$, we find $B = \dfrac{n_S}{\sum_{i=0}^4  \dfrac{A^i}{i!}\binom{4}{i}}$. Using condition $(2)$ we can then find the value of $A$. In summary, we find the values of $A$ and $B$ with:
\begin{equation}
    \def\arraystretch{1.7}
    \left\{
    \begin{array}{ll}
        \dfrac{\sum_{i=0}^4  i\dfrac{A^i}{i!}\binom{4}{i}}{\sum_{i=0}^4  \dfrac{A^i}{i!}\binom{4}{i}} = 4p & \\
        \dfrac{B}{n_S} = \dfrac{1}{\sum_{i=0}^4  \dfrac{A^i}{i!}\binom{4}{i}}
    \end{array}
    \right.
    \label{SolLagrange}
\end{equation}
This also shows that the normalization by $n_i$ is natural to lead to a probability distribution. We solve these two equations numerically to calculate the probability distribution $P_i = \frac{n_i}{n_S}$.
\section{Derivation of the proportions of $i$-loops}

We define $i$-loops as structures in which two stars are connected by $i$
 distinct chains. A $1$-loop is then a single strand between two stars, while $i$-loops with $i\geq 2$ are network defects. 
We first show the case where we only consider $2$-loops, and then extend our analysis to $i\geq 2$. We derive expressions for the proportions of $i$-loops by considering that these proportions should result in the maximization of the entropy associated with the topology of the network.
When only considering $2$-loops, we can consider that the entropy associated with the presence of loops results from the number of accessible microstates. As for the entropy associated with the connectivity of the stars, we assume that $p$ is fixed. We also assume that loops can only occur with formed bonds. This implies that $i$-loops can only form with stars that have $i$ or more connections. 
For $n_2$ loops, we can write the number of microstates as the number of ways to choose and pair $2n_2$ bonds. Assuming a total number of bonds formed $n_b$, the number of ways to choose $2n_2$ bonds is simply $\binom{n_b}{2n_2}$. The number of ways to pair these $2n_2$ bonds can be found by induction to be $\dfrac{(2n_2)!}{2^{n_2}n_2!}$. This result is known as the number of perfect matchings in graph theory, and proofs can be found in textboo. However, we give here a derivation as we will use the same idea for the cases $i=3$ and $i=4$.

To show this, we can first note that this expression is valid for $n_2$ = 1: we have then $\dfrac{(2n_2)!}{2^{n_2}n_2!} = 1$, and there is indeed only one way to pair 2 bonds. 
Let us now consider that this expression is valid for a given $n_2$, and consider the case with $2(n_2+1)$ bonds to pair. Let us pick a bond and call it the starting bond. There are then $2n_2+1$ bonds to choose from to create a first pair. Once this step is achieved, there are now $2n_2$ bonds left, and we can use the induction hypothesis. We find that there are $\left( 2n_2+1 \right) \times \dfrac{(2n_2)!}{2^{n_2}n_2!} = \dfrac{(2n_2+2)(2n_2+1)}{2(n_2+1)} \dfrac{(2n_2)!}{2^{n_2}n_2!} = \dfrac{(2(n_2+1))!}{2^{n_2+1}(n_2+1)!}$ ways to pair $2\left(n_2+1\right)$ bonds, confirming the induction hypothesis. One can notice that all possible pairs are taken into account only once with this reasoning: all possible pairs including the starting bond are counted only once by construction, and all possible pairs not including the starting bond are counted exactly once by the induction hypothesis. 

With this reasoning, we have counted the number of states associated with loops for a given number of loops $n_2$. However, we also have to take into account that creating loops will result in a decrease in the number of possible graphs: creating a 2-loop constrains one of the two bonds that are participating in the loop. This will lead to a decrease in the entropy associated with the number of possible graphs. We thus consider that the number of loops that maximizes the number of accessible microstates is determined by a competition between the number of ways to choose and pair bonds, and the associated decrease in the number of degrees of freedom of the total graph. 

To model the decrease in entropy associated with the number of possible graphs when creating loops, we consider the dominant term in Eq. \ref{eqSWMFin}: $S \approx \dfrac{1}{2}\sum_{i=0}^{4} in_i \times \log \left( \sum_{i=0}^{4} in_i\right)$. This approximation is justified in the thermodynamic limit, where $\sum_{i=0}^{4} in_i \gg 1$. For each 2-loop created, one bond is constrained. Hence, for $n_2$ 2-loops created, $n_2$ bonds are constrained. Using the fact that $\sum_{i=0}^{4} in_i = 2n_b$, we approximate the decrease in entropy associated with the number of possible graphs when creating loops as $\Delta S = n_2 \log (2n_2)$. 


We now combine the three terms mentioned above to calculate the entropy associated with loops: the number of ways to choose $2n_2$ bonds, the number of ways to pair them into $n_2$ loops, and the decrease in the entropy associated with the number of possible graphs. Using Stirling's approximation, we can write the resulting entropy as 
\begin{equation}
    S = k_B \left[ n_b\log n_b - (n_b-2n_2)\log (n_b-2n_2)-n_2 \log n_2 - n_2 (1+\log 2) - n_2 \log (2n_2)\right]
    \label{S2loop}
\end{equation}

We find the value of $n_2$ that maximizes this expression by finding the solution of $\frac{\partial S}{\partial n_2} = 0$. This yields:
\begin{equation}
    \dfrac{n_2}{n_b} = \dfrac{1}{2+2\sqrt{e}} \approx 0.19
    \label{2loopstResult}
\end{equation}
Importantly, the number of bonds $n_b$ in this equation is the number of bonds created by stars that have at least two connections: $n_b = \frac{1}{2} \left( 2P_2+3P_3+4P_4\right)$

Next, we have to consider the case of loops involving more than 2 bonds. We will restrict our analysis to loops involving up to 4 bonds, as we mainly used 4-arm star precursors, which can only form up to 4-loops. This derivation could be adapted to the case of stars with a higher functionality. 

We use the same reasoning to determine the number of microstates associated with loops of higher order: we calculate the number of ways to choose the bonds participating in loops, then count the number of ways to associate these bonds into loops, and finally consider that loops are indistinguishable. A difficulty arises when considering the number of bonds that are available to create loops. i-loops can form with stars that have more than $i$ bonds. However, when considering different types of loops, one has to take into account that the number of bonds available for one type of loop will be influenced by the number of loops of other types. For instance, when considering 2-loops only, we considered that the number of bonds available is $\frac{1}{2} \left( 2P_2+3P_3+4P_4\right)$. But when considering 3-loops and 4-loops as well, we have to take into account that some of the stars with 3 or 4 bonds will exhibit 2-loops. This leads to a decrease in the number of available bonds to form loops. 

In order to take this into account, we first solve the simple case where only one type of loop is considered. More precisely, what is the proportion of 3-loops that is entropically favorable when only 3-loops are considered, and what is the proportion of 4-loops that is entropically favorable when only 4-loops are considered. We treat each of these cases in a way similar to the approach developed above for the case where only 2-loops are considered. 

For the case of 3-loops, we note $n_b$ the number of bonds available to form loops and $n_3$ the number of 3-loops. We consider that, similarly to the case of 2-loops, the number of microstates comes from 3 contributions: 
\begin{enumerate}
    \item The number of ways to choose $3 n_3$ bonds among $n_b$ possible
    \item The number of ways to combine these $3 n_3$ bonds in $n_3$ 3-loops
    \item The decrease in entropy associated with the number of possible graphs
\end{enumerate}
The first contribution is given by the binomial coefficient $\binom{n_b}{3n_3}$. The second contribution is given by $\dfrac{(3n_3)!}{3!^{n_3}n_3!}$. This can be proven by induction with a similar reasoning as shown above for the case of 2-loops. Finally, for each loop created, 2 bonds are constrained, so we write the loss of entropy as $2 \times n_3 \log (2n_3)$. Combining all three terms, we write the entropy associated with the number of 3-loops as 
\begin{equation}
    \label{eqS3loops}
    S = k_B \left[ n_b\log n_b - (n_b-3n_3)\log (n_b-3n_3)-n_3 \log n_3 - n_3 (2+\log 6) - 2n_3 \log (2n_3) \right]
\end{equation}

We find the value of $n_3$ that maximizes this expression by finding the solution of $\frac{\partial S}{\partial n_3} = 0$. This yields:
\begin{equation}
    \dfrac{n_3}{n_b} = \dfrac{1}{3+\exp{\frac{2+\log 24}{3}}}
    \label{3loopstResult}
\end{equation}

For the case of 4-loops, we consider the same three contributions to the entropy associated with loops: 
\begin{enumerate}
    \item The number of ways to choose $4 n_4$ bonds among $n_b$ possible
    \item The number of ways to combine these $4 n_4$ bonds in $n_4$ 4-loops
    \item The decrease in entropy associated with the number of possible graphs
\end{enumerate}

The first contribution is given by the binomial coefficient $\binom{n_b}{4n_4}$. The second contribution is given by $\dfrac{(4n_4)!}{4!^{n_4}n_4!}$. This can be proven by induction with a similar reasoning as shown above for the case of 2-loops. Finally, for each loop created, 3 bonds are constrained, so we write the loss of entropy as $3 \times n_4 \log (2n_4)$. Combining all three terms, we write the entropy associated with the number of 3-loops as 
\begin{equation}
    \label{eqS4loops}
    S = k_B \left[ n_b\log n_b - (n_b-4n_4)\log (n_b-4n_4)-n_4 \log n_4 - n_4 (3+\log 4!) - 3n_4 \log (2n_4) \right]
\end{equation}

We find the value of $n_4$ that maximizes this expression by finding the solution of $\frac{\partial S}{\partial n_4} = 0$. This yields:
\begin{equation}
    \dfrac{n_4}{n_b} = \dfrac{1}{4+\exp{\frac{3+\log 192}{4}}}
    \label{4loopstResult}
\end{equation}

Having derived the fractions of $i$-loops for $i = $2, 3, or 4, for the case where only one type of loops is present, we now derive the fractions of loops for the case where all three types are present. The main difficulty lies in considering that the fraction of loops of one type influences the number of bonds available to form loops of a different type. In the derivation above, we considered that the number of bonds available to form $i$-loops is the number of bonds formed by stars with $i$ or more connections. In the case where several types of loops are present, the number of bonds available becomes the number of stars with $i$ connections not involved in a loop of a different type, resulting in a decrease in the number of bonds available, and hence in a decrease in the number of loops of each type compared to the case where only one type of loops is considered. 

To model this, we can use an iterative approach. The first step of the iteration is to consider that the fractions of loops derived above in the case where only one type of loop is present are a first approximation. We can now update the number of bonds available to form each type of loop by subtracting the number of bonds occupied by loops of another type. This is the second step of the iteration. More precisely, we note $n_i$ the number of $i$-loops. Then the number of bonds available to form 2-loops will be 
\begin{equation*}
    2P_2 + 3P_3 + 4P_4 - 3n_3 - 4n_4
\end{equation*}
This expression is also an approximation, as the number of loops $n_3$ and $n_4$ are overestimated. We will show below how to use this approximation to move toward a more accurate estimation of the number of loops. 

For the case of 3- and 4-loops, an additional subtlety arises. Let us consider for this the case of 3-loops. The number of bonds available to form loops is influenced by the number of 2-loops and 4-loops. However, some of the 2-loops form on stars that have only 2 connections. These loops thus do not decrease the number of bonds available to form 3-loops. Hence, we have to take into account only the number of 2-loops that form on stars with 3 or more connections. To calculate the number of such loops, we assume that the formation of a loop on 2 bonds of a star is not influenced by the number of bonds available on this star. We justify this assumption by considering that it is consistent with the maximization of information entropy: the greatest number of microstates possible is achieved when loops can form with equal probability on all stars with more than two connections regardless on the total number of connections. With this assumption, we can write the number of 2-loops that form on stars with 3 or more connections as 
\begin{equation}
n_2 \dfrac{3P_3+4P_4}{2P_2+3P_3+4P_4}
\end{equation}
The updated number of bonds available to form 3-loops, after the new iteration, then becomes
\begin{equation}
3P_3+4P_4-4n_4-2n_2\dfrac{3P_3+4P_4}{2P_2+3P_3+4P_4}
\end{equation}
Using a similar reasoning, the number of bonds available to form 4-loops becomes
\begin{equation}
4P_4-3n_3\dfrac{4P_4}{3P_3+4P_4}-2n_2\dfrac{4P_4}{2P_2+3P_3+4P_4}
\end{equation}
We can now use the relations derived previously to express the number of $i$-loops as a function of the number of bonds available, and consider the updated number of bonds available after the first iteration. Then the updated number of $i$-loops $n'_i$ are: 
\begin{equation}
    \arraycolsep=1.4pt\def\arraystretch{2.2}
    \left\{
    \begin{array}{ll}
        n'_2 = & \dfrac{1}{}\left( 2P_2+3P_3+4P_4 - 3n_3 - 4n_4 \right) \\
        n'_3 = & \dfrac{1}{K}\left( 3P_3+4P_4 - 2n_2\dfrac{3P_3+4P_4}{2P_2+3P_3+4P_4} - 4n_4 \right) \\
        n'_4 = & \dfrac{1}{K}\left( 4P_4-3n_3\dfrac{4P_4}{3P_3+4P_4}-2n_2\dfrac{4P_4}{2P_2+3P_3+4P_4} \right)
    \end{array}
    \right.
    \label{EqNi'Ni}
\end{equation}

We now consider that the correct proportion of loops in the network when taking into account different types of loops (here 2-, 3-, and 4-loops) corresponds to the steady-state of this iterative process. Indeed, the key idea of this process is to update the number of bonds available to determine a more accurate number of loops. If the number of loops does not change with this iterative process, this suggests that 

Following this reasoning, the number of loops $n_2$, $n_3$, and $n_4$ can thus be determined by solving the following system of linear equations: 

\begin{equation}
    \arraycolsep=1.4pt\def\arraystretch{2.2}
    \left\{
    \begin{array}{ll}
        n_2 = & \dfrac{1}{K_2}\left( 2P_2+3P_3+4P_4 - 3n_3 - 4n_4 \right) \\
        n_3 = & \dfrac{1}{K_3}\left( 3P_3+4P_4 - 2n_2\dfrac{3P_3+4P_4}{2P_2+3P_3+4P_4} - 4n_4 \right) \\
        n_4 = & \dfrac{1}{K_4}\left( 4P_4-3n_3\dfrac{4P_4}{3P_3+4P_4}-2n_2\dfrac{4P_4}{2P_2+3P_3+4P_4} \right)
    \end{array}
    \right.
    \label{EqNiIter}
\end{equation}
where $K_2$, $K_3$, and $K_4$ are the proportionality coefficient relating $n_b$ to $n_2$, $n_3$, $n_4$, respectively, and were determined in Eqs. \ref{2loopstResult}, \ref{3loopstResult}, \ref{4loopstResult}. 

We solve this system of equations to arrive at the final expression for the number of loops as a function of the quantities $P_2$, $P_3$, $P_4$: 

\begin{equation}
\arraycolsep=1.4pt\def\arraystretch{2.2}
    \left\{
    \begin{array}{ll}
    n_2 = &\frac{(2 P_2+3 P_3+4 P_4) (2 K_3 P_2+3 K_3 P_3+4 K_3 P_4-9 P_3-12 P_4) (K_4 (2
   P_2+3 P_3+4 P_4)-16 P_4)}{K_2 (2 P_2+3 P_3+4 P_4) (K_3 K_4 (2 P_2+3 P_3+4 P_4)-48 P_4)-32
   P_4 (2 (K_3-3) P_2+(K_3-6) (3 P_3+4 P_4))-6 K_4 (3 P_3+4 P_4) (2 P_2+3 P_3+4 P_4)}\\
   
   n_3 = &\frac{(K_2-2) (3 K_4 P_3+4 K_4 P_4-16 P_4) (2 P_2+3 P_3+4 P_4)^2}{K_2 (2 P_2+3 P_3+4 P_4) (K_3
   K_4 (2 P_2+3 P_3+4 P_4)-48 P_4)-32 P_4 (2 (K_3-3) P_2+(K_3-6) (3 P_3+4 P_4))-6 K_4 (3 P_3+4 P_4)
   (2 P_2+3 P_3+4 P_4)}\\
   
   n_4 = &\frac{4 (K_2-2) P_4 (2 P_2+3 P_3+4 P_4) (2 K_3 P_2+3 K_3 P_3+4 K_3
   P_4-9 P_3-12 P_4)}{K_2 (2 P_2+3 P_3+4 P_4) (K_3 K_4 (2 P_2+3 P_3+4 P_4)-48 P_4)-32 P_4 (2
   (K_3-3) P_2+(K_3-6) (3 P_3+4 P_4))-6 K_4 (3 P_3+4 P_4) (2 P_2+3 P_3+4 P_4)}
    \end{array}
    \right.
    \label{EqNiFinal}
\end{equation}


\section{Derivation of the model for $G'$}

We start from the phantom network model. This model assumes that the network is tree-like and that all chains are connected to the boundaries of the network via an infinite tree. The core idea of this model is to calculate the elastic behavior of this infinite tree. For this, it is possible to show that the tree can be modeled as a single polymer chain with length $K_\infty$ having the same elastic modulus. This is done using two properties of the elasticity of polymer chains: 
\begin{itemize}
    \item Two chains in series are elastically equivalent to a single chain, and the length of the single chain is the sum of the length of the two chains;
    \item Two chains in parallel are elastically equivalent to a single chain, and the inverse of the length of the single chain is the sum of the inverses of the lengths of the two chains
\end{itemize}

Using these two properties, one can derive the equivalent length of a perfect network with degree of polymerization between cross-links $N$ and functionality $f$: $K_\infty = N\frac{1}{f-2}$. Then, the length $N_\infty$ of a part of the network comprised of a single strand between two cross-links is the sum of this length and the lengths of the trees on both sides: 
\begin{equation}
    N_\infty = N+2K_\infty = N\dfrac{f}{f-2}
\end{equation}
Hence, each strand can be modeled as an equivalent chain of length $N\frac{f}{f-2}$ connected on both sides to the boundaries of the network. This chain will deform affinely with respect to the global deformation, so one can then use the affine network model to calculate the elastic modulus: 
\begin{equation}
    G' = \nu kT \dfrac{N}{N_\infty} = \nu kT \left(1-\dfrac{2}{f}\right)
    \label{GpPh}
\end{equation}

Our approach here is to calculate $N_\infty$ using a mean-field approximation, for a network with various strand lengths between cross-links and various functionalities. With a mean-field approximation, we have 
\begin{equation}
    N_\infty = \langle N\dfrac{f}{f-2} \rangle
\end{equation}
where $N$ and $f$ are now treated as random variables. We will show later that $N$ and $f$ are independent. In this case, one can write
\begin{equation}
    N_\infty = \langle N\rangle \langle \dfrac{f}{f-2} \rangle
\end{equation}

Let us first tackle the case of $\langle \dfrac{f}{f-2} \rangle$. Cross-links are stars with 3 or more connections to the infinite network. We note $P_i$ the probability for a star to have $i$ connections to the infinite network. Then we write 
\begin{equation}
    \langle \dfrac{f}{f-2} \rangle = \dfrac{\sum_{i\geq 3}\dfrac{i}{i-2}P_i}{\sum_{i\geq 3}P_i}
    \label{fav}
\end{equation}
We normalize by $\sum_{i>3}P_i$ because the calculation of $\langle \dfrac{f}{f-2} \rangle$ should be done only on stars that are cross-links. 

Let us now calculate the average length between cross-links $\langle N\rangle$. 
First, we have to consider that stars with two connections to the infinite network create long strands. Let us consider the total length of strands between cross-links. Each star with 2 or more connections to the infinite network contributes $N\times i/2$ to this total length, where $i$ is the number of arms connected to the infinite network. Hence, the total length of the strands between cross-links will be $\mu N\sum_{i\geq 2} \frac{i}{2}P_i$ where $\mu$ is the total number of stars. To calculate the total number of strands, we considered that each cross-link with functionality $f$
 creates $f/2$ strands. The total number of strands is then $\mu \sum_{i\geq 3} \frac{i}{2}P_i$. The average length between cross-links can then be found by the ratio of these two quantities: 
 \begin{equation}
     \langle N \rangle = \dfrac{\mu N\sum_{i\geq 2} \frac{i}{2}P_i}{\mu \sum_{i\geq 3} \frac{i}{2}P_i} = N\dfrac{\sum_{i\geq 2} \frac{i}{2}P_i}{\sum_{i\geq 3} \frac{i}{2}P_i}
     \label{Nav1}
 \end{equation}

From this, we can show that  

 Finally, we have to calculate the effective length between cross-links, by taking into account that loops can be modeled as shorter strands. An $i$-loop of strands of length $N$ can be modeled as a single chain of length $N/i$. In order to calculate the effective length of strands, we thus need to derive an expression for the average number of loops in a strand between cross-links. We assume that the probability of a loop is not correlated with the length of a strand. The probabilities of an $i$-loop are then global parameters noted $n_i$, and the average number of $i$-loops among $k$
 chains in series can be understood from the average of a binomial law: $k n_i$. 
Using eq.\ref{Nav1} we get the average number of $i$-loops between cross-links: 
\begin{equation}
    n_i \dfrac{\sum_{i\geq 2} \frac{i}{2}P_i}{\sum_{i\geq 3} \frac{i}{2}P_i}
\end{equation} 
An $i$-loop results in a decrease of the effective length of the chain by $N-N/i$. From this we get our final expression for the average effective length between cross-links: 
\begin{equation}
    \langle N \rangle = N\dfrac{\sum_{i\geq 2} \frac{i}{2}P_i}{\sum_{i\geq 3} \frac{i}{2}P_i} \left( 1-\sum_{i\geq 2}\dfrac{i-1}{i}n_i \right)
     \label{Nav2}
\end{equation}
In the case of a network formed by 4-arm stars, we can rewrite all sums as:
\begin{equation}
    \langle N \rangle = N\dfrac{P_2+1.5P_3+2P_4}{1.5P_3+2P_4} \left( 1-\dfrac{1}{2}n_2-\dfrac{2}{3}n_3 \right)
     \label{Navf}
\end{equation}
We can now see that the random variables $N$ and $f$ are independent, as required to obtain $\langle N \dfrac{f}{f-2} \rangle = \langle N\rangle \langle \dfrac{f}{f-2} \rangle$. $f$ is fully determined by the ration $\frac{P_3}{P_4}$ as shown by eq.\ref{fav}, while $N$ is fully determined by the ratios $\frac{P_3}{P_4}$ and $\frac{P_2}{P_4}$.

Finally, we need an expression for the number density of elastically effective strands $\nu$. This has been derived previously by several groups \cite{BrunoM,Parada2018} as 
\begin{equation}
    \nu = \mu \left( 1.5P_3+2P_4 \right)
    \label{nu}
\end{equation}

Combining eqs.\ref{GpPh}, \ref{fav}, \ref{Navf}, and \ref{nu}, we arrive at our final expression for the elastic modulus: 
\begin{equation}
    G' = \mu \left( 1.5P_3+2P_4 \right) kT \dfrac{1.5P_3+2P_4}{P_2+1.5P_3+2P_4} \dfrac{1}{1-\dfrac{1}{2}n_2-\dfrac{2}{3}n_3}\dfrac{P_3+P_4}{3P_3+2P_4}
\end{equation}

\section{Formation of networks based on disulfide bridges}

4-arm PEG stars with molecular weight 10kDa and end-functionalized with succinimidyl glutarate were purchased from JenKem USA. Cystamine dihydrochloride, 2-((2-aminoethyl)dithio)ethan-1-ol (abbreviated in the rest of the text as 2ADO), N,N-Diisopropylethylamine (DIPEA), and 1,8-Diazabicyclo[5.4.0]undec-7-ene (DBU) were purchased from Sigma-Aldrich. 
The networks were formed by linking PEG stars with cystamine. $p$ was controlled by the ratio of Cystamine to 2ADO. The gelation proceeded in a solution of 92.5 v/v\% dry dimethylsulfoxide (DMSO) and 7.5 v/v\% dry acetonitrile (ACN). Stock solutions of 20mM cystamine in dry DMSO and 40 mM 2ADO in a mixture of dry DMSO and dry ACN were prepared. The proportions of DMSO and ACN in the stock solution of 2ADO were adjusted so that their concentrations in the final mix was 92.5 v/v\% dry dimethylsulfoxide (DMSO) and 7.5 v/v\% dry acetonitrile (ACN). For each value of $p$, a solution with the corresponding ratio of Cystamine and 2ADO was prepared just before mixing with the PEG for each sample. The volume of solution added to the PEG stars was adjusted to arrive to a final PEG concentration of 10 wt\%, corresponsing to a concentration of succinimidyl glutarate of 40 mM. Following mixing with PEG, the solution was vortexed and 0.22 $\mu$L of DIPEA were added. The solution was then vortexed again and 50 $\mu$L were pipetted onto glass slides with an 8mm rubber spacer. Another glass slide was then added on top, and a 1 kg weight was added on top of the glass slide. The reaction was then left to proceed for three hours. 

\subsection{Measurement of the elastic properties of the networks}

Following gelation, the formed networks were removed from the glass slide and placed on a rheometer. After a 30s equilibration time, a frequency sweep from 100 rad/s to 0.1 rad/s was measured. The total duration of the measurement did not exceed 5 mn. Following this, the gel was removed from the rheometer and placed again on the same glass slide using the same rubber spacer. Then, 0.2 $\mu$L of DBU was added, and the gel was covered with a glass slide and a 1 kg weight in the same fashion as in the first step and the reaction was left to proceed for 12 hours. 

Following this, the gel was placed on the rheometer again. After a 30s equilibration time, a frequency sweep from 100 rad/s to 0.01 rad/s was measured. 

\section{Calculation of the final $p$}

Although no bond is lost during disulfide metathesis, the proportion of bonds formed $p$ may change. This is due to the formation of a new species: free chains in solution, analog to small molecules. The formation of such chains is illustrated on Fig. [ref]: when the disulfide bonds belonging to two dangling chains exchange via metathesis, the result can either be two other dangling chains, or one fully connected chains and one free chain. This free chain is copmosed of a disulfide bridge, but has no end connected to a polymer star. It can then also exchange with other disulfide bridges via metathesis. 

This effect leads to a change in $p$ when the metathesis reaction is allowed to occur. It is possible to understand this qualitatively by considering that dangling chains are characterized by a pendant -OH group. When metathesis is allowed to occur and free chains are formed, -OH groups distribute between dangling chains and free chains, therefore decreasing the number of dangling chains. This effect thus leads to an increase in $p$.

We propose here to relate the proportion of chains connected at both ends before and after the metathesis reaction is allowed to occur. The proportion of chains attached at both ends is defined as $p$ in our work. We define $p_0$ the proportion of chains bound at both ends before the metathesis reaction is allowed to occur, and $p_f$ after. 

We start by examining all possible metathesis reactions. We note C for chains that are connected to stars at both ends, D for dangling chains, and F for free chains. The possible reactions and their outcomes are summarized in the table below: 
\begin{table}
\arraycolsep=3pt
    \centering
    \begin{tabular}{cl}
        Reactants & Products\\
        C + C & C + C\\
        C + D & C + D\\
        C + F & D + D\\
        D + D & D + D or C + F\\
        D + F & D + F\\
        F + F & F + F\\
    \end{tabular}
    \caption{All possible metathesis reactions and the associated outcomes. C, D, and F refer to chains that are connected to stars at both ends, connected to a star at only one end, and not connected to a star at both ends, respectively.}
    \label{tab:Metareact}
\end{table}

From this, we can see that only the reactions D + D and C + F lead to a change in the composition of the system. For all other reactions, the reactants and products are the same. Assuming that metathesis occurs at the same rate $k$ for all reactions, we can write the rates of change of each species. We also assume that both outcomes of the metathesis reaction D + D are equally probable.  We will now assimilate species to their concentrations, such that C, D, F are the concentrations of connected, dangling, and free chains. From the table above, we get:
\begin{equation*}
\arraycolsep=1.4pt\def\arraystretch{2}
\left\{ 
\begin{array}{ll}
    \dfrac{\partial C}{\partial t} = & \dfrac{1}{2}k D^2 - kFC \\
    \dfrac{\partial D}{\partial t} = & -k D^2 + 2 kFC \\
    \dfrac{\partial F}{\partial t} = & \dfrac{1}{2}k D^2 - kFC \\
\end{array}
\right. 
\end{equation*}

We assume that equilibrium is reached when the rate of change of the concentrations is 0: 
\begin{equation*}
    \dfrac{\partial C}{\partial t} = \dfrac{\partial D}{\partial t} = \dfrac{\partial F}{\partial t} = 0
\end{equation*}
While this leads to three equations, we can easily notice that all three equations are the same, with the general form 
\begin{equation}
    FC-D^2 = 0
    \label{eqMeta1}
\end{equation}
We need two additional equations relating the concentrations C, D, F, in order to characterize the equilibrium. We first use the conservation of the number of disulfide bridges. If T is the total concentration of disulfide bridges in the system, then 
\begin{equation}
    C+D+F = T
    \label{eqMeta2}
\end{equation}
We also use the conservation of the -OH groups on pendant chains and write P their total concentration. There is one such group per dangling chain and two groups per free chain, yielding 
\begin{equation}
    D+2F = P
    \label{eqMeta3}
\end{equation}
We can now solve the system of equations described by eqs. \ref{eqMeta1}, \ref{eqMeta2}, \ref{eqMeta3} for C. The solution is: 
\begin{equation}
   C = \dfrac{1}{2}\left(-P + 3 T - \sqrt{-P^2 + 2 P T + T^2}\right)
   \label{eqMetaFin-1}
\end{equation}
We can now relate this result to the parameters $p_0$ and $p_f$. From the definition of C and P, we have $p_f = \frac{C}{T}$ and $1-p_0 = \frac{P}{T}$. Replacing C and P in eq. \ref{eqMetaFin-1} yields
\begin{equation}
    p_f = 1+\dfrac{p_0}{2}-\dfrac{1}{2}\sqrt{2-p_0^2}
    \label{eqMetaFin}
\end{equation}
This equation was used to ensure that the final proportion of chains connected at both ends corresponds to the expected value.

\subsection{}
\subsubsection{}

\bibliography{biblio}

@article{Nishi2014,
author = {Nishi, Kengo and Fujii, Kenta and Katsumoto, Yukiteru and Sakai, Takamasa and Shibayama, Mitsuhiro},
doi = {10.1021/ma500662j},
issn = {15205835},
journal = {Macromolecules},
month = {may},
number = {10},
pages = {3274--3281},
publisher = {American Chemical Society},
title = {{Kinetic aspect on gelation mechanism of tetra-peg hydrogel}},
url = {https://pubs.acs.org/sharingguidelines},
volume = {47},
year = {2014}
}

@article{Matsunaga2009a,
author = {Matsunaga, Takuro and Sakai, Takamasa and Akagi, Yuki and Chung, Ung-il and Shibayama, Mitsuhiro},
doi = {10.1021/ma901013q},
issn = {0024-9297},
journal = {Macromolecules},
month = {aug},
number = {16},
pages = {6245--6252},
publisher = {American Chemical Society},
title = {{SANS and SLS Studies on Tetra-Arm PEG Gels in As-Prepared and Swollen States}},
url = {https://pubs.acs.org/sharingguidelines https://pubs.acs.org/doi/10.1021/ma901013q},
volume = {42},
year = {2009}
}

@article{Sheridan2012,
author = {Sheridan, Richard J. and Bowman, Christopher N.},
doi = {10.1021/MA301329U/SUPPL_FILE/MA301329U_SI_001.PDF},
issn = {00249297},
journal = {Macromolecules},
month = {sep},
number = {18},
pages = {7634--7641},
publisher = {American Chemical Society},
title = {{A simple relationship relating linear viscoelastic properties and chemical structure in a model Diels-Alder polymer network}},
url = {https://pubs.acs.org/doi/full/10.1021/ma301329u},
volume = {45},
year = {2012}
}

@article{Semenov1998,
author = {Semenov, Alexander N. and Rubinstein, Michael},
doi = {10.1021/MA970616H/ASSET/IMAGES/MEDIUM/MA970616HE00083.GIF},
issn = {00249297},
journal = {Macromolecules},
month = {feb},
number = {4},
pages = {1373--1385},
publisher = {American Chemical Society},
title = {{Thermoreversible gelation in solutions of associative polymers. 1. Statics}},
url = {https://pubs.acs.org/doi/full/10.1021/ma970616h},
volume = {31},
year = {1998}
}

@article{Rubinstein1998,
author = {Rubinstein, Michael and Semenov, Alexander N},
doi = {10.1021/MA970617},
journal = {Macromolecules},
title = {{Thermoreversible Gelation in Solutions of Associating Polymers. 2. Linear Dynamics}},
url = {https://pubs.acs.org/journals/mamobx/index.html},
year = {1998}
}

@article{Brooks2018,
   author = {William L.A. Brooks and Christopher C. Deng and Brent S. Sumerlin},
   doi = {10.1021/ACSOMEGA.8B02999/ASSET/IMAGES/LARGE/AO-2018-029992_0001.JPEG},
   issn = {24701343},
   issue = {12},
   journal = {ACS Omega},
   month = {12},
   pages = {17863-17870},
   publisher = {American Chemical Society},
   title = {Structure-Reactivity Relationships in Boronic Acid-Diol Complexation},
   volume = {3},
   url = {https://pubs.acs.org/doi/full/10.1021/acsomega.8b02999},
   year = {2018},
}

@article{BrunoBE,
Author = {Marco-Dufort, B. and Tibbitt, M. W.},
Title = {Design of moldable hydrogels for biomedical applications using dynamic
   covalent boronic esters},
Journal = {MATERIALS TODAY CHEMISTRY},
Year = {2019},
Volume = {12},
Pages = {16-33},
Month = {JUN},
DOI = {10.1016/j.mtchem.2018.12.001},
ISSN = {2468-5194},
ResearcherID-Numbers = {Tibbitt, Mark W./A-4320-2019},
ORCID-Numbers = {Marco-Dufort, Bruno/0000-0001-9098-9964
   Tibbitt, Mark W./0000-0002-4917-7187},
Unique-ID = {WOS:000470901900003},
}

@book{SakaiBook,
author = {Sakai, Takamasa},
publisher = {John Wiley \& Sons, Ltd},
isbn = {9783527346547},
title = {Physics of Polymer Gels},
doi = {https://doi.org/10.1002/9783527346547},
url = {https://onlinelibrary.wiley.com/doi/abs/10.1002/9783527346547},
eprint = {https://onlinelibrary.wiley.com/doi/pdf/10.1002/9783527346547},
year = {2020}
}

@article{GelsreviewElastKey,
author = {Creton, Costantino},
title = {50th Anniversary Perspective: Networks and Gels: Soft but Dynamic and Tough},
journal = {Macromolecules},
volume = {50},
number = {21},
pages = {8297-8316},
year = {2017},
doi = {10.1021/acs.macromol.7b01698},

URL = { 
    
        https://doi.org/10.1021/acs.macromol.7b01698
    
    

},
eprint = { 
    
        https://doi.org/10.1021/acs.macromol.7b01698
    
    

}

}

@article{BrunoM,
author = {Marco-Dufort, Bruno and Iten, Ramon and Tibbitt, Mark W.},
title = {Linking Molecular Behavior to Macroscopic Properties in Ideal Dynamic Covalent Networks},
journal = {Journal of the American Chemical Society},
volume = {142},
number = {36},
pages = {15371-15385},
year = {2020},
doi = {10.1021/jacs.0c06192},
    note ={PMID: 32808783},

URL = { 
    
        https://doi.org/10.1021/jacs.0c06192
    
    

},
eprint = { 
    
        https://doi.org/10.1021/jacs.0c06192
    
    

}

}

@article{marco2022thermal,
  title={Thermal stabilization of diverse biologics using reversible hydrogels},
  author={Marco-Dufort, Bruno and Janczy, John R and Hu, Tianjing and L{\"u}tolf, Marco and Gatti, Francesco and Wolf, Morris and Woods, Alex and Tetter, Stephan and Sridhar, Balaji V and Tibbitt, Mark W},
  journal={Science advances},
  volume={8},
  number={31},
  pages={eabo0502},
  year={2022},
  publisher={American Association for the Advancement of Science}
}

@article{kloxin2013covalent,
  title={Covalent adaptable networks: smart, reconfigurable and responsive network systems},
  author={Kloxin, Christopher J and Bowman, Christopher N},
  journal={Chemical Society Reviews},
  volume={42},
  number={17},
  pages={7161--7173},
  year={2013},
  publisher={Royal Society of Chemistry}
}

@article{webber2022dynamic,
  title={Dynamic and reconfigurable materials from reversible network interactions},
  author={Webber, Matthew J and Tibbitt, Mark W},
  journal={Nature Reviews Materials},
  volume={7},
  number={7},
  pages={541--556},
  year={2022},
  publisher={Nature Publishing Group UK London}
}

@article{winne2019dynamic,
  title={Dynamic covalent chemistry in polymer networks: a mechanistic perspective},
  author={Winne, Johan M and Leibler, Ludwik and Du Prez, Filip E},
  journal={Polymer Chemistry},
  volume={10},
  number={45},
  pages={6091--6108},
  year={2019},
  publisher={Royal Society of Chemistry}
}

@article{craig2005,
author = {Yount, Wayne C. and Loveless, David M. and Craig, Stephen L.},
title = {Small-Molecule Dynamics and Mechanisms Underlying the Macroscopic Mechanical Properties of Coordinatively Cross-Linked Polymer Networks},
journal = {Journal of the American Chemical Society},
volume = {127},
number = {41},
pages = {14488-14496},
year = {2005},
doi = {10.1021/ja054298a},
    note ={PMID: 16218645},

URL = { 
    
        https://doi.org/10.1021/ja054298a
    
    

},
eprint = { 
    
        https://doi.org/10.1021/ja054298a
    
    

}

}

@article{Parada2018,
   author = {German Alberto Parada and Xuanhe Zhao},
   doi = {10.1039/c8sm00646f},
   issn = {17446848},
   issue = {25},
   journal = {Soft Matter},
   month = {6},
   pages = {5186-5196},
   pmid = {29780993},
   publisher = {Royal Society of Chemistry},
   title = {Ideal reversible polymer networks},
   volume = {14},
   url = {https://pubs.rsc.org/en/content/articlehtml/2018/sm/c8sm00646f https://pubs.rsc.org/en/content/articlelanding/2018/sm/c8sm00646f},
   year = {2018},
}

@article{Nishi2012,
   author = {Kengo Nishi and Masashi Chijiishi and Yukiteru Katsumoto and Toshio Nakao and Kenta Fujii and Ung Il Chung and Hiroshi Noguchi and Takamasa Sakai and Mitsuhiro Shibayama},
   doi = {10.1063/1.4769829/194834},
   issn = {00219606},
   issue = {22},
   journal = {Journal of Chemical Physics},
   month = {12},
   pages = {224903},
   pmid = {23249028},
   publisher = {AIP Publishing},
   title = {Rubber elasticity for incomplete polymer networks},
   volume = {137},
   url = {/aip/jcp/article/137/22/224903/194834/Rubber-elasticity-for-incomplete-polymer-networks},
   year = {2012},
}

@article{LangPreStretch,
   author = {Michael Lang},
   doi = {10.1021/ACSMACROLETT.8B00020/SUPPL_FILE/MZ8B00020_SI_003.TXT},
   issn = {21611653},
   issue = {5},
   journal = {ACS Macro Letters},
   month = {5},
   pages = {536-539},
   pmid = {35632927},
   publisher = {American Chemical Society},
   title = {Elasticity of Phantom Model Networks with Cyclic Defects},
   volume = {7},
   year = {2018},
}

@article{BrassartPreStretch,
   author = {Ghadeer Alamé and Laurence Brassart},
   doi = {10.1115/1.4048316},
   issn = {15289036},
   issue = {12},
   journal = {Journal of Applied Mechanics, Transactions ASME},
   month = {12},
   publisher = {American Society of Mechanical Engineers (ASME)},
   title = {Effect of topological defects on the elasticity of near-ideal polymer networks},
   volume = {87},
   url = {http://asmedigitalcollection.asme.org/appliedmechanics/article-pdf/87/12/121006/6566364/jam_87_12_121006.pdf},
   year = {2020},
}

@article{Miller1976,
   author = {Douglas R. Miller and Christopher W. Macosko},
   doi = {10.1021/MA60050A004/ASSET/MA60050A004.FP.PNG_V03},
   issn = {15205835},
   issue = {2},
   journal = {Macromolecules},
   month = {3},
   pages = {206-211},
   publisher = {American Chemical Society},
   title = {A New Derivation of Post Gel Properties of Network Polymers},
   volume = {9},
   url = {https://pubs.acs.org/doi/abs/10.1021/ma60050a004},
   year = {1976},
}

@article{lange2011connectivity,
  title={Connectivity and structural defects in model hydrogels: A combined proton NMR and Monte Carlo simulation study},
  author={Lange, Frank and Schwenke, Konrad and Kurakazu, Manami and Akagi, Yuki and Chung, Ung-il and Lang, Michael and Sommer, Jens-Uwe and Sakai, Takamasa and Saalwachter, Kay},
  journal={Macromolecules},
  volume={44},
  number={24},
  pages={9666--9674},
  year={2011},
  publisher={ACS Publications}
}

@article{lin2018topological,
  title={Topological structure of networks formed from symmetric four-arm precursors},
  author={Lin, Tzyy-Shyang and Wang, Rui and Johnson, Jeremiah A and Olsen, Bradley D},
  journal={Macromolecules},
  volume={51},
  number={3},
  pages={1224--1231},
  year={2018},
  publisher={ACS Publications}
}

@article{olsenNDS,
  title={Counting primary loops in polymer gels},
  author={Zhou, Huaxing and Woo, Jiyeon and Cok, Alexandra M and Wang, Muzhou and Olsen, Bradley D and Johnson, Jeremiah A},
  journal={Proceedings of the National Academy of Sciences},
  volume={109},
  number={47},
  pages={19119--19124},
  year={2012},
  publisher={National Academy of Sciences}
}

@article{nishi2017experimental,
  title={Experimental observation of two features unexpected from the classical theories of rubber elasticity},
  author={Nishi, Kengo and Fujii, Kenta and Chung, Ung-il and Shibayama, Mitsuhiro and Sakai, Takamasa},
  journal={Physical Review Letters},
  volume={119},
  number={26},
  pages={267801},
  year={2017},
  publisher={APS}
}

@article{rent,
  title={Quantifying the impact of molecular defects on polymer network elasticity},
  author={Zhong, Mingjiang and Wang, Rui and Kawamoto, Ken and Olsen, Bradley D and Johnson, Jeremiah A},
  journal={Science},
  volume={353},
  number={6305},
  pages={1264--1268},
  year={2016},
  publisher={American Association for the Advancement of Science}
}

@article{DBUSS,
  title={Dynamic disulfide metathesis induced by ultrasound},
  author={Fritze, Urs F and von Delius, Max},
  journal={Chemical Communications},
  volume={52},
  number={38},
  pages={6363--6366},
  year={2016},
  publisher={Royal Society of Chemistry}
}

@article{DBUSS2,
  title={Dynamic covalent chemistry of disulfides offers a highly efficient synthesis of diverse benzofused nitrogen- sulfur heterocycles in one pot},
  author={Zhu, Ning and Zhang, Fa and Liu, Gang},
  journal={Journal of Combinatorial Chemistry},
  volume={12},
  number={4},
  pages={531--540},
  year={2010},
  publisher={ACS Publications}
}

@inproceedings{rentPreStrain,
  title={Extending the phantom network theory to account for cooperative effect of defects},
  author={Lin, Tzyy-Shyang and Wang, Rui and Johnson, Jeremiah A and Olsen, Bradley D},
  booktitle={Macromolecular symposia},
  volume={385},
  pages={1900010},
  year={2019},
  organization={Wiley Online Library}
}

@incollection{cousin2025dynamic,
  title={Dynamic Biomaterials from Reversible Covalent Interactions},
  author={Cousin, Lucien and Tibbitt, Mark W},
  booktitle={Advanced Polymer Life Science},
  pages={279--326},
  year={2025},
  publisher={World Scientific}
}

@article{Zou2017,
author = {Zou, Weike and Dong, Jiante and Luo, Yingwu and Zhao, Qian and Xie, Tao and Zou, W K and Dong, J T and Luo, Y W and Zhao, Q and Xie, T},
doi = {10.1002/ADMA.201606100},
issn = {1521-4095},
journal = {Advanced Materials},
month = {apr},
number = {14},
pages = {1606100},
pmid = {28221707},
publisher = {John Wiley \& Sons, Ltd},
title = {{Dynamic Covalent Polymer Networks: from Old Chemistry to Modern Day Innovations}},
url = {https://onlinelibrary.wiley.com/doi/full/10.1002/adma.201606100 https://onlinelibrary.wiley.com/doi/abs/10.1002/adma.201606100 https://onlinelibrary.wiley.com/doi/10.1002/adma.201606100},
volume = {29},
year = {2017}
}

@article{VanZee2020,
author = {{Van Zee}, Nathan J. and Nicola{\"{y}}, Renaud},
doi = {10.1016/J.PROGPOLYMSCI.2020.101233},
issn = {0079-6700},
journal = {Progress in Polymer Science},
month = {may},
pages = {101233},
publisher = {Pergamon},
title = {{Vitrimers: Permanently crosslinked polymers with dynamic network topology}},
volume = {104},
year = {2020}
}

@article{Zheng2021,
author = {Zheng, Ning and Xu, Yang and Zhao, Qian and Xie, Tao},
doi = {10.1021/acs.chemrev.0c00938},
title = {{Dynamic Covalent Polymer Networks: A Molecular Platform for Designing Functions beyond Chemical Recycling and Self-Healing}},
journal={Chemical Reviews},
url = {https://dx.doi.org/10.1021/acs.chemrev.0c00938},
year = {2021}
}

@article{Akagi2013,
author = {Akagi, Yuki and Gong, Jian Ping and Chung, Ung Il and Sakai, Takamasa},
doi = {10.1021/MA302270A/ASSET/IMAGES/LARGE/MA-2012-02270A_0007.JPEG},
issn = {00249297},
journal = {Macromolecules},
month = {feb},
number = {3},
pages = {1035--1040},
publisher = {American Chemical Society},
title = {{Transition between phantom and affine network model observed in polymer gels with controlled network structure}},
url = {https://pubs.acs.org/doi/full/10.1021/ma302270a},
volume = {46},
year = {2013}
}

@article{Zhang,
author = {Zhang, Vivian and Kang, Boyeong and Accardo, Joseph V and Kalow, Julia A},
doi = {10.1021/jacs.2c08104},
title = {{Structure--Reactivity--Property Relationships in Covalent Adaptable Networks}},
journal={Journal of the American Chemical Society},
year={2022},
url = {https://doi.org/10.1021/jacs.2c08104}
}

@article{Fujiyabu2019b,
author = {Fujiyabu, Takeshi and Yoshikawa, Yuki and Chung, Ung-Il and Sakai, Takamasa},
doi = {10.1080/14686996.2019.1618685},
journal = {Science and Technology of Advanced Materials},
title = {{Structure-property relationship of a model network containing solvent}},
url = {https://doi.org/10.1080/14686996.2019.1618685},
volume = {20},
year = {2019}
}

@article{Lin2018,
author = {Lin, Tzyy Shyang and Wang, Rui and Johnson, Jeremiah A. and Olsen, Bradley D.},
doi = {10.1021/ACS.MACROMOL.7B01829/ASSET/IMAGES/MA-2017-018295_M006.GIF},
issn = {15205835},
journal = {Macromolecules},
month = {feb},
number = {3},
pages = {1224--1231},
publisher = {American Chemical Society},
title = {{Topological Structure of Networks Formed from Symmetric Four-Arm Precursors}},
url = {https://pubs.acs.org/doi/full/10.1021/acs.macromol.7b01829},
volume = {51},
year = {2018}
}

@article{McKay1991,
author = {McKay, Brendan D. and Wormald, Nicholas C.},
doi = {10.1007/BF01275671/METRICS},
issn = {02099683},
journal = {Combinatorica},
month = {dec},
number = {4},
pages = {369--382},
publisher = {Springer-Verlag},
title = {{Asymptotic enumeration by degree sequence of graphs with degrees o(n1/2)}},
url = {https://link.springer.com/article/10.1007/BF01275671},
volume = {11},
year = {1991}
}

@misc{pietroDPD,
  author = {Miotti, Pietro and Cousin, Lucien and Tibbitt, Mark W. and  Pivkin, Igor V.},
  title = {Mesoscopic Modeling of Dynamic Tetra-PEG Hydrogel Networks},
  publisher = {arXiv},
  doi = {10.48550/arXiv.2603.17180},
  note = {Avalailable at https://arxiv.org/abs/2603.17180},
  year = {2026},
  copyright = {CC-BY}
}

\end{document}